\documentclass[12pt]{article}
\usepackage{jheppub} 
\usepackage{amsfonts,amsmath,amssymb,epsf,mathtools,dsfont}
\usepackage{amsmath,braket}
\usepackage{graphicx,hyperref,color}
\usepackage{pgfplots}
\usepackage{comment}
\hypersetup{
    bookmarks=true,         
    unicode=false,          
    pdftoolbar=true,        
    pdfmenubar=true,        
    linktocpage=true,       
    pdffitwindow=false,     
    pdfstartview={FitH},    
    pdfnewwindow=true,      
    colorlinks=true,       
    linkcolor=blue,          
    citecolor=blue,        
    filecolor=blue,      
    urlcolor=blue           
}

\numberwithin{equation}{section}									

\newcommand{\be}{\begin{equation}}
\newcommand{\ba}{\begin{eqnarray}}
\newcommand{\ea}{\end{eqnarray}}
\newcommand{\ee}{\end{equation}}

\newcommand{\s}{\sqrt}

\newcommand{\ti}{\tilde}
\newcommand{\ap}{\alpha}

\newcommand{\no}{\nonumber \\}
\newcommand{\la}{\langle}
\newcommand{\lb}{\rangle}
\newcommand{\bea}{\begin{eqnarray}}
\newcommand{\eea}{\end{eqnarray}}
\newcommand{\bes}{\begin{equation*}}
\newcommand{\beas}{\begin{eqnarray*}}
\newcommand{\eeas}{\end{eqnarray*}}
\newcommand{\bas}{\begin{array*}}
\newcommand{\eas}{\end{array*}}
\newcommand{\ees}{\end{equation*}}
\newcommand{\nn}{\nonumber}

\newcommand{\ep}{\epsilon}

\def\nn{\nonumber}
\def\inf{\infty}

\usepackage{amsmath}

\renewcommand{\Re}{\operatorname{Re}}
\renewcommand{\Im}{\operatorname{Im}}

\subheader{\today}
\title{\boldmath Entanglement Suppression Due to Black Hole Scattering}

\author[a]{Kazuki Doi,}
\author[a,b]{Tadashi Takayanagi}
\affiliation[a]{Center for Gravitational Physics and Quantum Information, Yukawa Institute for Theoretical Physics, Kyoto University,\\
	Kitashirakawa Oiwakecho, Sakyo-ku, Kyoto 606-8502, Japan}
\affiliation[b]{Inamori Research Institute for Science,\\
620 Suiginya-cho, Shimogyo-ku,Kyoto 600-8411 Japan}

\abstract{We consider the evolution of entanglement entropy in a two-dimensional conformal field theory with a holographic dual. Specifically, we are interested in a class of excited states produced by a combination of pure-state (local operator) and mixed-state local quenches. We employ a method that allows us to determine the full time evolution analytically. While a single insertion of a local operator gives rise to a logarithmic time profile of entanglement entropy relative to the vacuum, we find that this growth is heavily suppressed in the presence of a mixed-state quench, reducing it to a time-independent constant bump. The degree of suppression depends on the relative position of the quenches as well as the ratio of regularization parameters associated with the quenches. This work sheds light on the interesting properties of gravitational scattering involving black holes.}

\begin{document} 

\begin{flushright}
YITP-25-103\\
\end{flushright}
\maketitle
\flushbottom

\clearpage
\section{Introduction}\label{Sec:Intro}

Entanglement entropy provides a useful and universal probe of non-equilibrium dynamics of quantum many-body systems. In thermal equilibrium, the degrees of freedom in a given quantum system is described by a single quantity of thermal entropy as the system is translationally invariant and time-independent. Under generic non-equilibrium time evolution, entanglement entropy $S_A$, defined for various choices of subsystem $A$, can measure the degrees of freedom or the amount of quantum correlation in each subsystem as a function of time. This quantity is highly universal in that it can be defined in any quantum systems without relying on any special features as opposed to many other quantities such as correlation functions or Wilson loops.

In quantum field theories, locality leads to a well-known property called the area law \cite{Bombelli:1986rw,Srednicki:1993im}, which states that entanglement entropy is UV-divergent and its leading contribution is proportional to the area of the boundary of subsystem $A$. In two-dimensional conformal field theories (CFTs), this divergence is known to become logarithmic \cite{Holzhey:1994we,Calabrese:2004eu,Casini:2009sr,Calabrese:2009qy} as an exception to the area law. The divergent term is universal in that it is determined once we fix the field theory Lagrangian and subsystem $A$, and that it does not depend on what kind of excited states we consider. Therefore, the finite difference $\Delta S_A = S_A-S^{(0)}_A$, where $S^{(0)}_A$ is the entanglement entropy in the ground state, is a useful quantity from which we can understand the non-equilibrium dynamics of a quantum field theory \cite{Calabrese:2005in,Calabrese:2007mtj}. In holography \cite{Maldacena:1997re}, entanglement entropy can equivalently be computed as the area of extremal surfaces \cite{Ryu:2006bv,Ryu:2006ef,Hubeny:2007xt,Nishioka:2009un,Rangamani:2016dms}, and thus this finite shift can be interpreted as the change in extremal surface area under a deformation of the dual gravity background. 

One of the simplest setups of an excited state that has a clear holographic dual is the local operator quench\footnote{In this paper, we will sometimes refer to the local operator quench as `pure-state local quench' or simply `pure-state excitation', and these names will be used interchangeably.}, which was first introduced in \cite{Nozaki:2014hna,Nozaki:2014uaa} with its gravitational dual given in \cite{Nozaki:2013wia} in terms of a falling heavy particle in anti-de Sitter (AdS) space. CFT calculations reveal that the time evolution of entanglement entropy $\Delta S_A$ is sensitive to the chaotic nature of the CFT that we consider. In two-dimensional integrable CFTs, entanglement entropy increases by only a finite amount, given by the quantum dimension of the operator that was inserted to excite the system \cite{He:2014mwa,Nozaki:2014hna,Nozaki:2014uaa} (see also \cite{Caputa:2014eta,Guo:2015uwa,Caputa:2015tua,Chen:2015usa,Nozaki:2015mca,David:2016pzn,Caputa:2016yzn}). In the symmetric orbifold of a compactified scalar with irrational radius, it grows like $\sim \log\log t$ \cite{Caputa:2017tju}. On the other hand, for two-dimensional holographic CFTs \cite{Fitzpatrick:2014vua,Hartman:2014oaa}, which are expected to be maximally chaotic, it grows logarithmically $\sim\frac{c}{6}\log t$ when the subsystem is very large \cite{Asplund:2014coa} (see also \cite{Caputa:2014vaa,Kusuki:2017upd,Kusuki:2018wpa}). This behavior can be explained holographically by taking into account the backreactions of heavy particles in its gravitational dual \cite{Nozaki:2013wia,Caputa:2014vaa}. Local operator quenches in higher-dimensional CFTs have also recently been studied in \cite{Bianchi:2025fzs}. Another setup we can consider is the mixed-state local quench, which is a different class of local quench produced by a localized mixed state: a canonical distribution of all possible local primary operators and their descendants in a CFT \cite{Bhattacharyya:2019ifi}. The holographic dual of the mixed-state local quench is given as a localized black hole in the AdS geometry. Interestingly, the mixed-state local quench is locally equivalent to the local operator quench under a certain identification of their parameters, which was pointed out in \cite{Bhattacharyya:2019ifi} and will be examined more in the present paper. Moreover, a class of excitations which sit in the middle of local and global quenches has been worked out recently \cite{Goto:2021sqx,Goto:2023wai,Goto:2023yxb,Kudler-Flam:2023ahk,Mao:2024cnm,Miyata:2024gvr,Bai:2024azk}.

As a next step, to further extend our understanding of the dynamics of entanglement in CFTs, it would be intriguing to consider two coexisting local excitations. Via holography, this is dual to the dynamics of two interacting heavy particles in AdS gravity, which indeed involves richer dynamics. In the case of two-dimensional integrable CFTs, this was computed in \cite{Numasawa:2016kmo} and it was shown that the scattering of two local operators does not change the growth of entanglement entropy; the individual contribution of each local operator excitation simply adds up. On the other hand, for holographic CFTs, it was found via the light cone bootstrap approach \cite{Fitzpatrick:2014vua,Kusuki:2018wpa,Collier:2018exn} that the logarithmic growth of entanglement entropy is reduced in the case of a double local operator quench \cite{Kusuki:2019avm,Kusuki:2019gjs}. A setup involving two joining local quenches \cite{Calabrese:2007mtj,Ugajin:2013xxa,Shimaji:2018czt} was also studied along with its gravitational dual in \cite{Caputa:2019avh}, where a similar suppression in entanglement entropy was observed.

In this paper, we would like to further explore double local quenches in a two-dimensional holographic CFT by considering the setup where one of the two local excitations is an excitation due to a heavy primary local operator and the other is a mixed-state excitation. This hybrid setup of a double local quench describes scattering of an entangled pair of quanta off a localized thermal gas. This is holographically dual to the scattering of a heavy particle off a localized black hole in AdS. As we shall see, introducing a mixed-state local quench enables us to employ the heavy-heavy-light-light conformal block approximation \cite{Fitzpatrick:2014vua} and explicitly evaluate the time evolution of entanglement entropy under a double local quench.

This paper is organized as follows: in section \ref{Sec:HHLL}, we review the local operator quench and its gravitational dual. 
In section \ref{Sec:mixed}, we briefly review the mixed-state local quench and introduce our hybrid construction of a class of double local quenches. We present basic analytic computations of the entanglement entropy and energy density under this class of double local quenches. We will also discuss our results from the viewpoint of the duality between the local operator quench and the mixed-state local quench.
In section \ref{Sec:suppression}, we present computational results of our analysis and provide holographic interpretations. We will also discuss some of the difficulties that are encountered in our analysis and examine the validity of the approximation that is central to the analysis.
In section \ref{Sec:CD}, we summarize our findings and discuss future problems. In appendix \ref{ap:thermal}, we present an analysis of the thermal AdS phase. In appendix \ref{ap:pauli}, we discuss how our analysis can be extended to include the chemical potential as well.

\section{Single local operator quenches}\label{Sec:HHLL}
The main purpose of this paper is to study the time evolution of entanglement entropy in the presence of two different excitations in a two-dimensional holographic CFT: (1) a local operator (pure-state) excitation and (2) a mixed-state excitation. As we shall see in section \ref{Sec:mixed}, we can treat the latter geometrically and the doubly-excited state can be analyzed by performing a path integral of a single local operator quench on a geometry that is obtained by modifying a Euclidean plane. Before we can do that, however, we must first know how to perform this path integral on an ordinary Euclidean plane; this section will devoted to learning how to do just that --- by reviewing the calculation of entanglement entropy under a single local operator excitation, following \cite{Asplund:2014coa}. We shall see that the heavy-heavy-light-light (HHLL) approximation of the identity conformal block \cite{Fitzpatrick:2014vua} can be employed for holographic CFTs, and the results perfectly match with those from \cite{Nozaki:2013wia}. We will also provide a simple holographic interpretation of the HHLL calculation of entanglement entropy, which shall prove to be useful for later arguments.

\subsection{HHLL approximation of conformal blocks}\label{HHLL}

Consider a two-dimensional holographic CFT, in which we shall insert a spinless heavy ($h = \bar{h} \sim O(c)$) primary operator $\mathcal{O}$ at $x = 0$. This gives us a UV-regulated state 
(${\mathcal{N}}_{\mathcal{O}}$ is the overall normalization factor):
\be
{\mathcal{N}}_{\mathcal{O}} e^{-\delta H}\mathcal{O}(t)|0\lb,
\label{localOPD}
\ee
where $H$ is the CFT Hamiltonian and $\delta$ is the regularization parameter. Taking subsystem $A$ to be a single interval $[a,b]$, we are interested in computing $\Delta S_A$, the entanglement entropy of this excited state relative to the vacuum. Morally speaking, the operator insertion produces an entangled pair of excitations that travel in opposite directions at the speed of light. The location of the pair in relation to the subsystem is what determines $\Delta S_A$. Specifically, when both modes of the pair are inside/outside the subsystem, entanglement entropy is expected to vanish. On the other hand, when only one of the modes is inside the subsystem, there is entanglement between the subsystem and its complement, giving rise to nontrivial entanglement entropy $\Delta S_A \neq 0$. This physical interpretation of the local quench is often referred to as the quasiparticle picture. While the analytical result for this particular setup is already known, the purpose of this section is to review the method that is employed in the computation and to apply it on a setup that involves an additional excitation in the form of a mixed-state local operator insertion.

\begin{figure}
    \centering
    \includegraphics[width=0.7\linewidth]{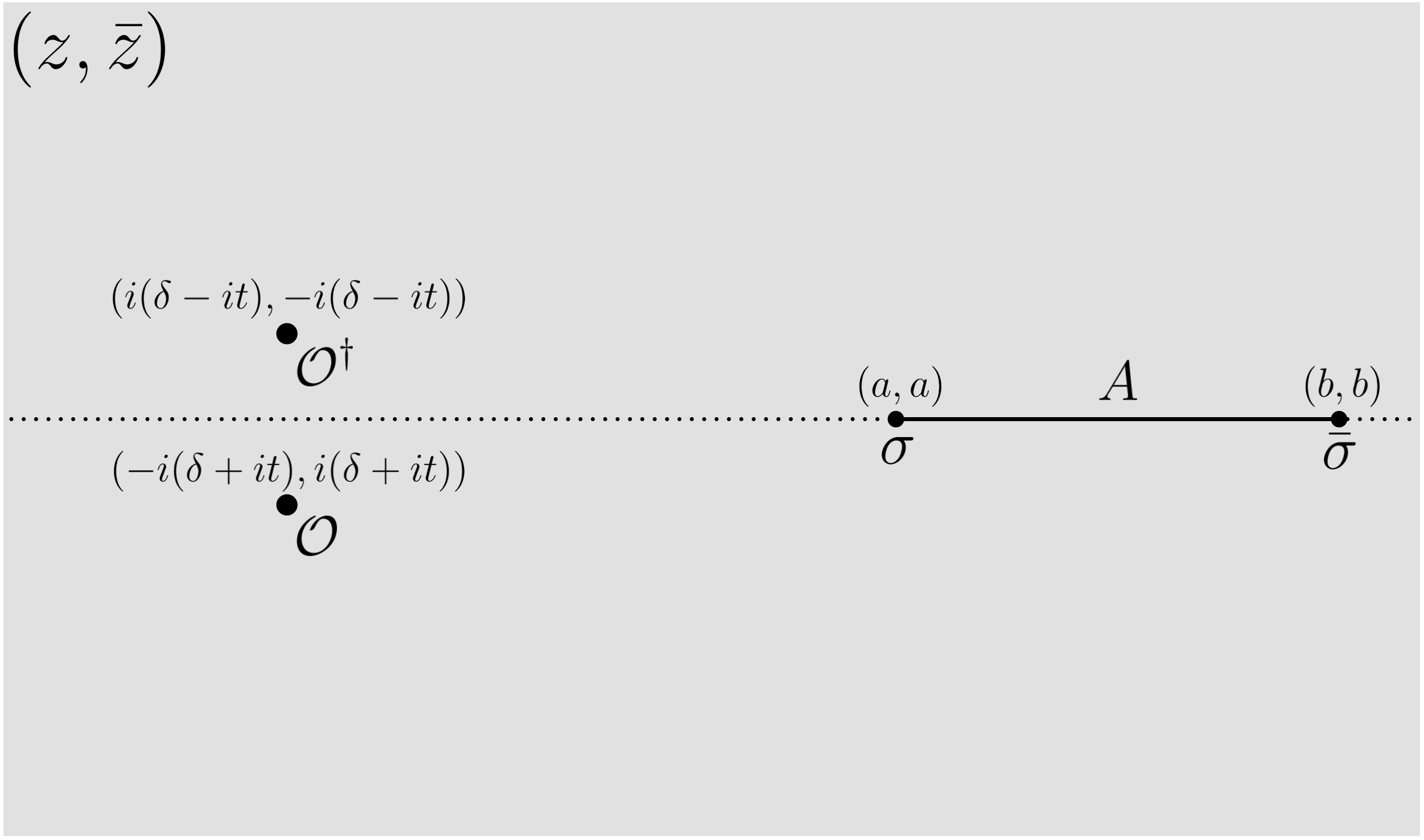}
    \caption{Setup for the single local operator quench.}
    \label{fig:singleexcitationsetup}
\end{figure}

The usual procedure of the replica trick is to first compute the $n$-th R\'enyi entropy, which we shall do by performing a Euclidean path integral. Passing on to the Euclidean plane (see figure \ref{fig:singleexcitationsetup}), the R\'enyi entropy relative to the vacuum is expressed as
\begin{align}
    \Delta S^{(n)}_A &= \frac{1}{1-n}\log\left(\frac{\langle{\mathcal{O}}^{\dagger\otimes n}(w_1,\bar{w}_1)\mathcal{O}^{\otimes n}(w_2,\bar{w}_2)\sigma_n(a) \bar{\sigma}_n(b)\rangle}{\langle{\mathcal{O}}^{\dagger\otimes n}(w_1,\bar{w}_1)\mathcal{O}^{\otimes n}(w_2,\bar{w}_2)\rangle \langle\sigma_n(a)\bar{\sigma}_n(b)\rangle}\right) \nonumber\\
    &= \frac{1}{1-n}\log\left(|z|^{4h_{\sigma_n }}G(z,\bar{z})\right),
\end{align}
where $\sigma_n, \bar{\sigma}_n$ are twist operators and the insertions of operators $\mathcal{O},\mathcal{O}^\dagger$ are located at
\begin{align}
    w_1 &= i(\delta-it), & \bar{w}_1 &= -i(\delta-it), \\
    w_2 &= -i(\delta+it), & \bar{w}_2 &= i(\delta+it),
\end{align}
remembering that $\delta$ is the regularization parameter from (\ref{localOPD}) that ensures we are considering normalizable states. Finally, $(z,\bar{z})$ are the cross ratios
\begin{align}
    z &= \frac{(a-b)(w_1-w_2)}{(a-w_1)(b-w_2)}, \\
    \bar{z} &= \frac{(a-b)(\bar{w}_1-\bar{w}_2)}{(a-\bar{w}_1)(b-\bar{w}_2)},
\end{align}
and $G(z,\bar{z})$, as a function of these cross ratios, is defined as
\begin{equation}
    G(z,\bar{z}) \equiv \frac{\langle{\mathcal{O}}^{\dagger\otimes n}(1)\mathcal{O}^{\otimes n}(\infty)\sigma_n (0)\bar{\sigma}_n(z,\bar{z})\rangle}{\langle{\mathcal{O}}^{\dagger\otimes n}(1)\mathcal{O}^{\otimes n}(\infty)\rangle},
\end{equation}
i.e. the normalized four-point function.

In a general CFT, the four-point function can be expressed as an expansion in terms of conformal blocks:
\begin{equation}
    G(z,\bar{z}) = \sum_p C_{\mathcal{O}^{\dagger\otimes n}\mathcal{O}^{\otimes n}}^p C_{\sigma_n\bar{\sigma}_n}^p \mathcal{F}(h_p,h_{\mathcal{O}^{\otimes n}},h_{\sigma_n}|z) \mathcal{F}(\bar{h}_p,\bar{h}_{\mathcal{O}^{\otimes n}},\bar{h}_{\sigma_n}|\bar{z}),
\end{equation}
where the sum runs over conformal families whose primary states $\mathcal{O}_p$ are of dimensions $(h_p,\bar{h}_p)$. When the CFT has a holographic dual, it is expected that the theory has a sparse spectrum of light operators $\Delta \ll c$ and a dense spectrum of heavy operators $\Delta \sim c$ corresponding to conical defects and black hole microstates. Under this assumption, the four-point function will be dominated by contributions from the identity block $\mathcal{O}_p = \mathds{1}$. The expression for the R\'enyi entropy simplifies to
\begin{equation}
    \Delta S_A^{(n)} = \frac{1}{1-n} \log\left(|z|^{4h_{\sigma_n}}|\mathcal{F}(0,h_{\mathcal{O}^{\otimes n}},h_{\sigma_n}|z)|^2\right).
\end{equation}
The problem is now reduced to computing the conformal block $\mathcal{F}(0,h_{\mathcal{O}^{\otimes n}},h_{\sigma_n}|z)$.

The conformal dimension of twist operators $h_{\sigma_n}=\bar{h}_{\sigma_n}=\frac{c}{24}(n-\frac{1}{n})$ is heavy in general, but becomes light in the limit of $n \to 1$. As we are interested in this limit, we can employ the HHLL approximation of conformal blocks.
\begin{equation}
    \mathcal{F}(0,h_{\mathcal{O}^{\otimes n}},h_{\sigma_n}|z) = \left(\frac{\alpha_H}{1-(1-z)^{\alpha_H}}\right)^{2h_{\sigma_n}} (1-z)^{-h_{\sigma_n}(1-\alpha_H)},
\end{equation}
where $\alpha_H = \sqrt{1-24h_{\mathcal{O}}/c}$. Making the substitution for $h_{\sigma_n}$ and taking the limit $n \to 1$, we arrive at the result
\begin{equation}\label{HHLLformula}
    \Delta S_A = \frac{c}{6}\log\left(\frac{1}{|z|^2}\left|\frac{1-(1-z)^{\alpha_H}}{\alpha_H}\right|^2 |1-z|^{1-\alpha_H}\right).
\end{equation}

\subsection{Computation of entanglement entropy}\label{singleinsertion}

Given (\ref{HHLLformula}), our focus shifts to the computation of cross ratios $(z,\bar{z})$, which is straightforward for this particular setup. They come out as
\begin{align}
    1-z &= \frac{(a-w_2)(b-w_1)}{(a-w_1)(b-w_2)} \nn\\ 
        &= \frac{(a-t+i\delta)(b-t-i\delta)}{(a-t-i\delta)(b-t+i\delta)}, \label{zevo} \\
    1-\bar{z} &= \frac{(a-\bar{w}_2)(b-\bar{w}_1)}{(a-\bar{w}_1)(b-\bar{w}_2)} \nn\\ 
        &= \frac{(a+t-i\delta)(b+t+i\delta)}{(a+t+i\delta)(b+t-i\delta)}. \label{zbevo}
\end{align}
Notice that both $1-z$ and $1-\bar{z}$ are of unit length, so let us define $\theta, \bar{\theta}$ in a way such that they satisfy
\begin{equation}
    1-z = e^{i\theta}, \qquad 1-\bar{z} = e^{-i\bar{\theta}}.
    \label{zphasew}
\end{equation}
Using this notation, we can greatly simplify (\ref{HHLLformula}) to give
\begin{equation}\label{HHLLsimple}
    \Delta S_A = \frac{c}{6}\log\left(\frac{\sin\left(\frac{\alpha_H \theta}{2}\right)\sin\left(\frac{\alpha_H \bar{\theta}}{2}\right)}{\alpha^2_H \sin\left(\frac{\theta}{2}\right)\sin\left(\frac{\bar{\theta}}{2}\right)}\right).
\end{equation}
Immediately, we can point out its resemblance to the known expression for the geodesic length in AdS$_3$ with a defect. We should also note that the allowed range of values for $\theta$, $\bar{\theta}$ should at least be restricted to $|\theta|, |\bar{\theta}|<2\pi$ so as to avoid $\Delta S_A$ being imaginary, remembering that $\alpha_H<1$.

We wish to work in the limit $\delta \ll 1$, so (\ref{zevo}) and (\ref{zbevo}) reduce to
\begin{align}
    1-z &= 1 + 2i\delta\left(\frac{1}{a-t} - \frac{1}{b-t}\right) + O(\delta^2), \\
    1-\bar{z} &= 1 - 2i\delta\left(\frac{1}{a+t} - \frac{1}{b+t}\right) + O(\delta^2).
\end{align}
In terms of $\theta$, $\bar{\theta}$, there is an ambiguity in the choice of branch
\begin{align}
    \theta &= 2\pi m + 2\delta\left(\frac{1}{a-t} - \frac{1}{b-t}\right) + O(\delta^2), \\
    \bar{\theta} &= 2\pi \bar{m} + 2\delta\left(\frac{1}{a+t} - \frac{1}{b+t}\right) + O(\delta^2),
\end{align}
where $|m|, |\bar{m}| \le 1$ (as $|\theta|, |\bar{\theta}|<2\pi$). The correct choice of branch will need to be determined on physical grounds, as both $1-z$ and $1-\bar{z}$ exhibit monodromic behavior depending on the position of the operator insertion relative to the subsystem.

\paragraph{Insertion outside subsystem.}
Let us first consider the case where the operator insertion lies outside the subsystem, $ab>0$. There are two possible configurations: $a,b>0$ (which we shall call `case I') and $a,b<0$ (`case III'). Without loss of generality, assume case I. For $t<a$, causality tells us that we should expect $\Delta S_A = 0$, which can only be achieved if $m=\bar{m}=0$. Indeed, starting from this branch and letting $z$, $\bar{z}$ evolve in time, we see that for $a<t<b$, $m=1$ while $\bar{m}=0$, which gives us finite $\Delta S_A$:
\begin{equation}\label{outside}
    \Delta S_A = \frac{c}{6}\log\left(\frac{\sin\left(\alpha_H \pi\right)(t-a)(b-t)}{\alpha_H \delta (b-a)}\right).
\end{equation}
For $t>b$, we are once again back at $m=\bar{m}=0$, giving us $\Delta S_A = 0$ as expected. In terms of the quasiparticle picture, while the left-moving mode is always outside the subsystem, the right-moving mode enters the subsystem at $t=a$ and leaves at $t=b$, matching the expectation that $\Delta S_A$ is non-vanishing for $a<t<b$.

\paragraph{Insertion inside subsystem.} Let us now consider the case where the operator insertion lies inside the subsystem, $a<0<b$ (`case II'). Once again, without loss of generality, assume $|a|<b$. The quasiparticle picture tells us that $\Delta S_A = 0$ when $t<|a|$ (as both modes are inside the subsystem) and $t>b$ (as both modes are outside the subsystem), which implies $m=\bar{m}=0$ for those times. However, there seemingly arises a problem of consistency when we impose those conditions. If we impose $m=\bar{m}=0$ for $t>b$ and let $z,\bar{z}$ evolve backwards in time, we see that for $|a|<t<b$, $m=1,\bar{m}=0$, and for $t<|a|$, $m=\bar{m}=1$. This gives us
\begin{equation}
    \Delta S_{A,1} =
    \begin{cases}
        \frac{c}{6}\log\left(\frac{\sin^2\left(\alpha_H \pi\right)(t^2-a^2)(b^2-t^2)}{\alpha_H^2 \delta^2 (b-a)^2}\right), & t<|a| \\
        \frac{c}{6}\log\left(\frac{\sin\left(\alpha_H \pi\right)(t-a)(b-t)}{\alpha_H \delta (b-a)}\right), & |a| < t < b  \\
        0, & t > b 
    \end{cases}
\end{equation}
where 1 in the subscript of $\Delta S_{A,1}$ implies we are considering a particular channel that fixes $(m,\bar{m})$ at some point in time. In order to impose $m=\bar{m}=0$ for $t<|a|$, we have to consider another channel by replacing $1-z \mapsto e^{-2\pi i}(1-z)$ and $1-\bar{z} \mapsto e^{2\pi i}(1-\bar{z})$, which corresponds to a shift of $(m,\bar{m})\mapsto(m-1,\bar{m}-1)$. The second channel gives
\begin{equation}
    \Delta S_{A,2} =
    \begin{cases}
        0, & t<|a| \\
        \frac{c}{6}\log\left(\frac{\sin\left(\alpha_H \pi\right)(t+a)(b+t)}{\alpha_H \delta (b-a)}\right), & |a| < t < b  \\
        \frac{c}{6}\log\left(\frac{\sin^2\left(\alpha_H \pi\right)(t^2-a^2)(b^2-t^2)}{\alpha_H^2 \delta^2 (b-a)^2}\right). & t > b 
    \end{cases}
\end{equation}
Among these two candidate channels, the correct choice at any given time should always minimize the entanglement entropy, so
\begin{equation}
    \Delta S_A = \min_{i} \Delta S_{A,i},
\end{equation}
where $i$ labels the channel. Applying this prescription, we finally obtain
\begin{equation}\label{inside}
    \Delta S_A =
    \begin{cases}
        0, & t<|a|, t>b \\
        \frac{c}{6}\log\left(\frac{\sin\left(\alpha_H \pi\right)(t+a)(b+t)}{\alpha_H \delta (b-a)}\right), & |a| < t < \sqrt{-ab}  \\
        \frac{c}{6}\log\left(\frac{\sin\left(\alpha_H \pi\right)(t-a)(b-t)}{\alpha_H \delta (b-a)}\right). & \sqrt{-ab} < t < b 
    \end{cases}
\end{equation}

Notably for both cases, i.e. regardless of the position of the insertion, in the limit of a large subsystem and late time $|a| \ll t \ll b$, we get
\begin{equation}
    \Delta S_A = \frac{c}{6}\log\left(\frac{\sin\left(\alpha_H \pi\right)t}{\alpha_H \delta}\right).\label{logwlq}
\end{equation}
For later convenience, let us call the channel that imposes $m=\bar{m}=0$ at late times `channel 1' and the other channel `channel 2'. By comparing the functional form of (\ref{outside}) and (\ref{inside}), we can see that channel 1 is the only dominant channel and channel 2 is irrelevant when the insertion is outside the subsystem.

\paragraph{General behavior of $\theta, \bar{\theta}$.}
Compiling these results and using the symmetry of the system, we arrive at the following general behavior of $\theta, \bar{\theta}$.
\begin{itemize}
    \item When the right-propagating mode of the entangled pair reaches
    \begin{itemize}
        \item the left endpoint $a$ of the subsystem: $(m,\bar{m})\mapsto(m+1,\bar{m})$,
        \item the right endpoint $b$ of the subsystem: $(m,\bar{m})\mapsto(m-1,\bar{m})$.
    \end{itemize}
    \item When the left-propagating mode of the entangled pair reaches
    \begin{itemize}
        \item the right endpoint $b$ of the subsystem: $(m,\bar{m})\mapsto(m,\bar{m}+1)$,
        \item the left endpoint $a$ of the subsystem: $(m,\bar{m})\mapsto(m,\bar{m}-1)$.
    \end{itemize}
\end{itemize}
The different channels are simply there to set different initial conditions for $(m,\bar{m})$. For case I, we only see monodromic behavior from $1-z$, whereas for case III, we only see monodromic behavior from $1-\bar{z}$. In either of these cases, $(m,\bar{m})$ at $t=0$ and at large $t$ are always the same, which is why there is only one dominant channel. For case II, we see monodromic behavior from both $1-z$ and $1-\bar{z}$, but they revolve once and do not revolve back to $1-z=1-\bar{z}=1$. So $(m,\bar{m})$ at $t=0$ and at large $t$ are always different, which necessitates contribution from a secondary channel.

Another related observation that we can make from (\ref{HHLLsimple}) is that only the variables that exhibit monodromy matter. Provided that $\theta,\bar{\theta}$ can be treated as small angles at $m=0$ or $\bar{m}=0$, we see that $\Delta S_A$ assumes a nonzero value only when $m\neq0$ or $\bar{m}\neq0$. So in case I (III), we should only care about the behavior of $z$ ($\bar{z}$); we do not have to carry out explicit calculations to find out the value $\bar{\theta}$ ($\theta$) assumes. Obviously, from our previous arguments, case II is a lot more subtle and both $z$ and $\bar{z}$ have to be carefully tracked. We shall see in section \ref{Sec:mixed} that in the double excitation setup, the assumption of $\theta,\bar{\theta}$ being small at $m=0$ or $\bar{m}=0$ starts to fail in certain case II configurations, especially when the regulators for the two excitations become comparable to each other in size.

\subsection{Holographic interpretation}\label{sec:holg}

Consider the AdS$_3/$CFT$_2$ duality in global coordinates:
\ba
ds^2=-\cosh^2\rho dt^2+d\rho^2+\sinh^2\rho d\phi^2.
\ea

The geodesic distance $D_{12}$ between two points $(\rho_1,t_1,\phi_1)$ and $(\rho_2,t_2,\phi_2)$ (this interval defines subsystem $A$) is given by the formula
\ba
\cosh D_{12}=\cos(t_1-t_2)\cosh\rho_1\cosh\rho_2-\cos(\phi_2-\phi_1)\sinh \rho_1 \sinh\rho_2.
\ea
By writing $t_2-t_1=\Delta t$ and $\phi_2-\phi_1=\Delta \phi$, we find that when the two points are both at the AdS boundary $\rho_1=\rho_2=\rho_\infty\to \infty$,
\ba
D_{12}\simeq  2\rho_\infty+\log\left(\sin \left(\frac{\Delta \phi+\Delta t}{2}\right)\sin \left(\frac{\Delta \phi-\Delta t}{2}\right)\right).
\ea

Now we shall introduce a deficit angle such that $\phi\sim \phi+2\pi\ap$ with $0<\ap<1$. By rescaling the coordinates $\phi=\ap\ti{\phi}, t=\ap\ti{t}$ and shifting the cutoff to 
$e^{\ti{\rho}_\infty}=e^{\rho_\infty}\ap$, we obtain 
\ba
D_{12}\simeq 2\ti{\rho}_\infty+\log\left(\frac{\sin \left(\frac{\ap}{2}\left(\Delta \ti{\phi}+\Delta \ti{t}\right)\right)\sin \left(\frac{\ap}{2}\left(\Delta \ti{\phi}-\Delta \ti{t}\right)\right)}{\ap^2}\right). \label{EEdefi}
\ea
Now, if we introduce angles $\theta, \bar{\theta}$, defined as
\begin{equation}\label{thephs}
\theta=\Delta\ti{\phi}-\Delta\ti{t}, \qquad \bar{\theta}=\Delta\ti{\phi}+\Delta\ti{t},
\end{equation}
then these agree with the angles in (\ref{zphasew}).
Note that $\Delta\ti\phi$ is meaningful only in the range $0<\Delta\ti\phi<2\pi$ as $\ti\phi\sim\ti\phi+2\pi$. The entanglement entropy comes out as 
\begin{align}\label{holoEE}
S_A(\alpha)&=\frac{1}{4G_N}\min\left[D_{12}(\Delta\ti{\phi},\Delta \ti{t}),\ 
D_{12}(2\pi-\Delta\ti{\phi},\Delta \ti{t})\right]\nonumber\\
&=\frac{1}{4G_N}\min\left[D_{12}(\theta,\bar{\theta}),\ 
D_{12}(2\pi-\bar{\theta},2\pi-\theta)\right]\nonumber\\
&=\frac{1}{4G_N}\min\left[D_{12}(\theta,\bar{\theta}),\ 
D_{12}(\theta-2\pi,\bar{\theta}-2\pi)\right],
\end{align}
where the second geodesic is obtained by taking the subsystem that wraps around the other side of the cylinder $\Delta\ti\phi\mapsto2\pi-\Delta\ti\phi$. From (\ref{holoEE}), it is evident that taking the second geodesic is equivalent to choosing channel 2, replacing $\theta\mapsto\theta-2\pi$, $\bar{\theta}\mapsto\bar{\theta}-2\pi$ and hence $(m,\bar{m})\mapsto(m-1,\bar{m}-1)$. It follows that $\Delta S_A = S_A(\alpha) - S_A(1)$ reproduces the result (\ref{HHLLsimple}) obtained from the HHLL calculation of holographic CFTs.

\section{Formulation of mixed-state local quenches}\label{Sec:mixed}

Now we move on to the main analysis of this paper. We are ultimately interested in doubly-excited states, where one of the excitations is a local operator excitation, thoroughly analyzed in the previous section, and the other is a mixed-state excitation, first introduced in \cite{Bhattacharyya:2019ifi}. As will be explained later, mixed-state excitations can be realized by opening up and gluing two holes in the Euclidean plane on which we are performing a path integral. This allows us to treat these quenches geometrically. Our final objective is to perform the same analysis of a single local operator excitation on this modified geometry.

\begin{figure}[h]
    \centering
    \includegraphics[width=0.7\linewidth]{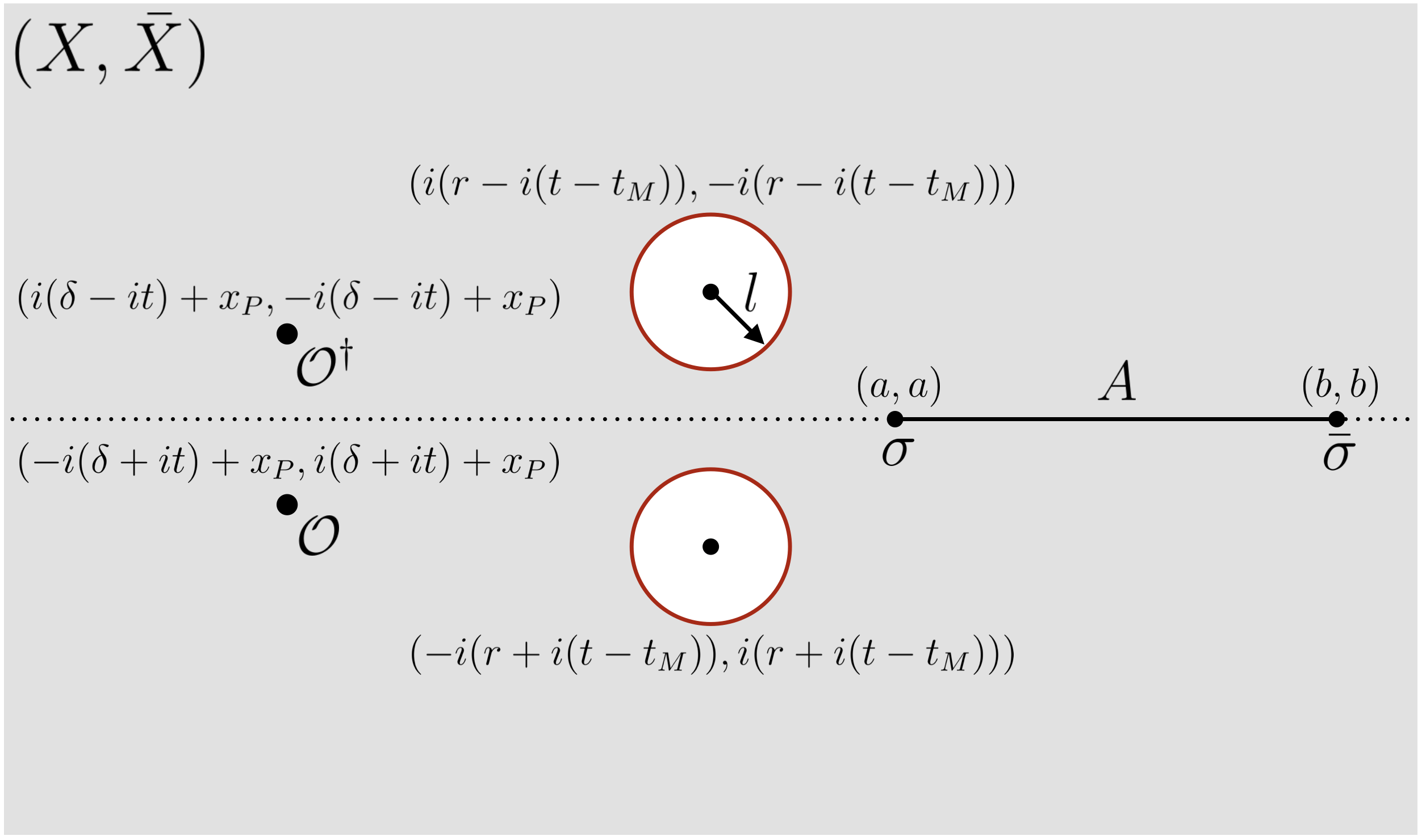}
    \caption{Our setup of interest depicted on the Euclidean plane. The gluing of two holes produces a mixed-state excitation, while the insertion of operator $\cal O$ gives rise to a local excitation. We measure the entanglement entropy $S_A$ for the interval $A$.}
    \label{fig:setup}
\end{figure}

\subsection{Mixed-state local quench}\label{mixedsingle}

Let us first briefly review the mixed-state local quench, which is a class of local excitations of the form \cite{Bhattacharyya:2019ifi}:
\begin{equation}\label{mixedexcitation}
\rho(t)=\sum_{i} \frac{e^{-\beta (\Delta_i - \frac{c}{12})}}{Z(\beta)}|\mathcal{O}_i(t)\lb \la \mathcal{O}_i(t)|,
\end{equation}
where $\beta$ is the inverse temperature of the local excitation and the sum runs over all states (primary states and their descendants) of the CFT. As in (\ref{localOPD}), states $|\mathcal{O}_i(t)\lb$ are UV-regulated such that
\be
|\mathcal{O}_i(t)\lb={\cal N}_i e^{-sH}\mathcal{O}_i(t)|0\lb,
\ee
where ${\cal N}_i$ is the normalization constant and $s$ is the regularization parameter. So we can see that mixed-state excitations are defined by two parameters $\beta, s$. We shall insert this excitation at $(x,t)=(0,0)$ and track the evolution of entanglement between subsystem $A$, defined by the interval $0<a<x<b$, and its complement.

This state (\ref{mixedexcitation}) can be realized in the Euclidean path integral by creating finite-sized holes of radius $l$ and identifying them (see figure \ref{fig:setup}, but ignore the operator insertions $\mathcal{O},\mathcal{O}^\dagger$ and variable $t_M$, at least for now) \cite{Bhattacharyya:2019ifi}. The locations of the holes (their centers) are given in terms of the following complex coordinates:
\begin{align}
    X_1 &= i(r-it), & \bar{X}_1 &= -i(r-it), \\
    X_2 &= -i(r+it), & \bar{X}_2 &= i(r+it),
\end{align}
where $r$ is yet another regularization parameter. These newly-introduced parameters $(r,l)$ are related to $(\beta,s)$ by
\begin{align}
    r &= \frac{s}{\tanh(\beta/2)}, \\
    l &= \frac{s}{\sinh(\beta/2)},
\end{align}
or equivalently,
\begin{align}
    \beta &= \log\left(\frac{r+\sqrt{r^2-l^2}}{r-\sqrt{r^2-l^2}}\right), \\
    s &= \sqrt{r^2-l^2}.
\end{align}
Any mixed-state excitations can be defined by specifying either $(r,l)$ or $(\beta,s)$, but since we are interested in fixing $\beta$ in this paper, we will primarily work with the latter.

\begin{figure}
    \centering
    \includegraphics[width=0.6\linewidth]{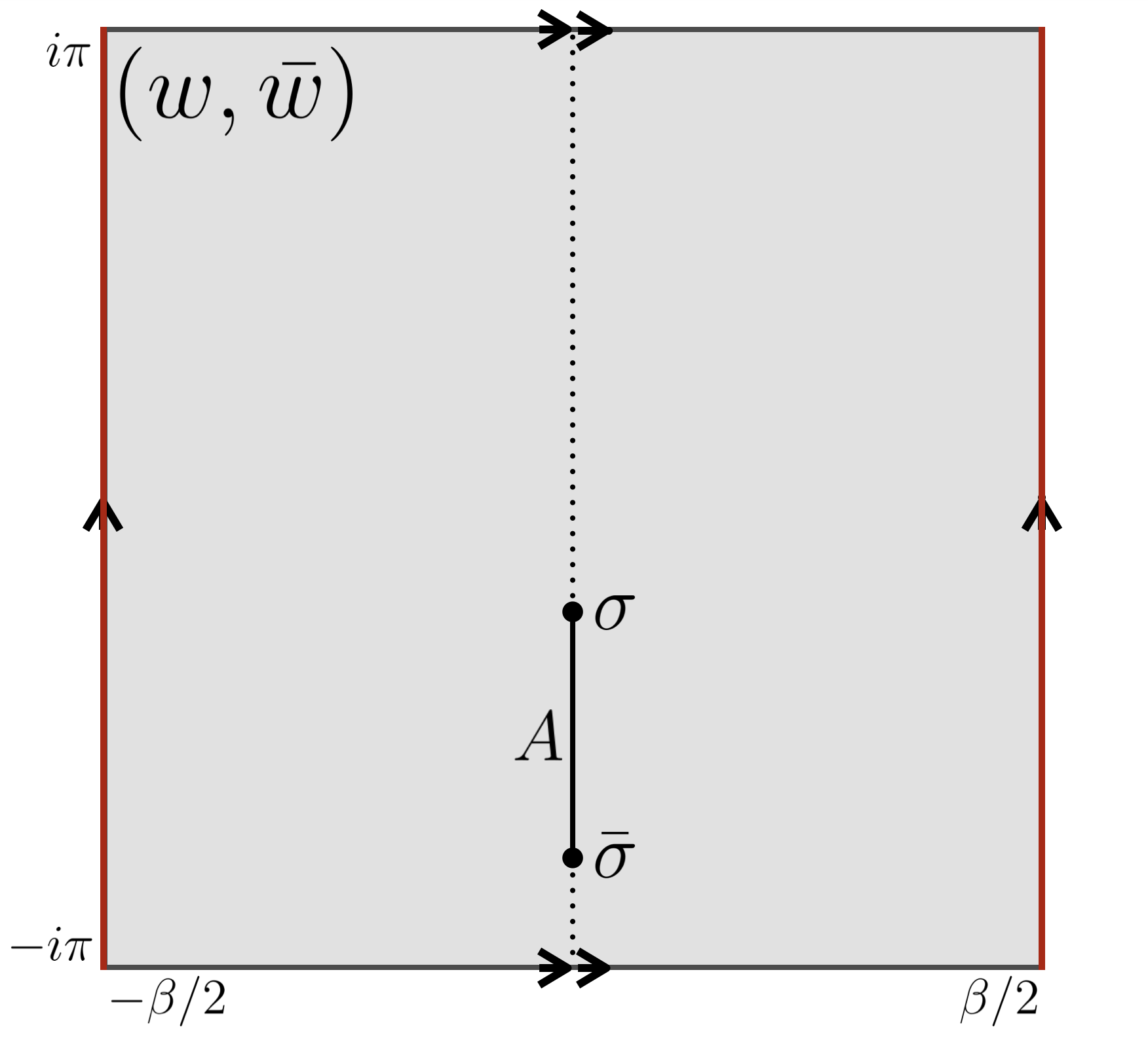}
    \caption{The original two-hole geometry $(X,\bar{X})$ is conformally transformed into a torus $(w,\bar{w})$. The two holes in figure \ref{fig:setup} are mapped onto the vertical boundaries of the strip, as indicated by the red lines. They are to be identified. The horizontal boundaries are also identified as the spatial direction is compactified.}
    \label{fig:w_coord}
\end{figure}

This geometry can be conformally mapped onto a torus $(w,\bar{w})$ with a temporal period of $\beta$ via the following transformation:
\begin{align}
    w &= \log\left(-\frac{X-t+is}{X-t-is}\right), \label{torustransf} \\
    \bar{w} &= \log\left(-\frac{\bar{X}+t-is}{\bar{X}+t+is}\right). \label{torustransfb}
\end{align}
The conformally-transformed geometry is shown in figure \ref{fig:w_coord}.

The problem of computing correlation functions under this excitation reduces to computing them on a torus, which usually does not help a great deal given that they depend on the details of the CFT of interest. However, for holographic CFTs, the partition function on a torus undergoes a sharp phase transition at $\beta = 2\pi$, and on either side of the phase transition the torus can effectively be treated as an infinitely long cylinder. In the high-temperature phase $\beta < 2\pi$, the spatial direction is decompactified and the resulting cylinder is dual to the BTZ black hole. In the low-temperature phase $\beta > 2\pi$, the temporal direction is decompactified and the resulting cylinder is dual to thermal AdS space. Luckily, we do know the behavior of correlation functions on a cylinder. In particular, we are interested (for now) in computing the two-point function of twist operators $\langle\sigma_n(a)\bar{\sigma}_n(b)\rangle$.

Let us focus on the high-temperature phase, $\beta < 2\pi$. The two-point function of twist operators in a finite-temperature CFT is known to be
\begin{equation}\label{finitetemp}
\langle\sigma_n(w_a,\bar{w}_a)\bar{\sigma}_n(w_b,\bar{w}_b)\rangle = \left(\frac{\beta^2}{\pi^2 \epsilon^2}\sinh\left(\frac{\pi(\Theta_b-\Theta_a)}{\beta}\right)\sinh\left(\frac{\pi(\bar{\Theta}_b-\bar{\Theta}_a)}{\beta}\right)\right)^{-\frac{c}{12}(n-\frac{1}{n})},
\end{equation}
where $(\Theta,\bar{\Theta})=(\Im w,-\Im \bar{w})$ (for real $X$, $w$ will come out to be purely imaginary). $\epsilon$ is also introduced as a UV cutoff (lattice spacing). Now, in order to go back to the original geometry with holes, one simply has to apply the following relation:
\begin{align}\label{striptotwohole}
\langle\sigma_n(a)\bar{\sigma}_n(b)\rangle &= \left|\frac{dw}{dX}\right|_{X=a}^{\frac{c}{12}(n-\frac{1}{n})}\left|\frac{dw}{dX}\right|_{X=b}^{\frac{c}{12}(n-\frac{1}{n})}\langle\sigma_n(w_a,\bar{w}_a)\bar{\sigma}_n(w_b,\bar{w}_b)\rangle \nonumber\\
&= \left(\frac{\beta^2}{\pi^2 \epsilon_a \epsilon_b}\sinh\left(\frac{\pi(\Theta_b-\Theta_a)}{\beta}\right)\sinh\left(\frac{\pi(\bar{\Theta}_b-\bar{\Theta}_a)}{\beta}\right)\right)^{-\frac{c}{12}(n-\frac{1}{n})},
\end{align}
where
\begin{equation}
    \epsilon_x = \frac{2s}{\sqrt{((x+t)^2+s^2)((x-t)^2+s^2)}}\epsilon.
\end{equation}
Finally, we apply our usual replica trick to compute the entanglement entropy,
\begin{align}\label{BTZ_EE}
S_A^\text{mixed} &= - \lim_{n \to 1}\left[\frac{\partial}{\partial n}\log\langle\sigma_n(a)\bar{\sigma}_n(b)\rangle\right] \nonumber\\
&= \frac{c}{6}\log\left(\frac{\beta^2}{\pi^2 \epsilon_a \epsilon_b}\sinh\left(\frac{\pi(\Theta_b-\Theta_a)}{\beta}\right)\sinh\left(\frac{\pi(\bar{\Theta}_b-\bar{\Theta}_a)}{\beta}\right)\right).
\end{align}
Working in the limit of a large subsystem $a \ll b$ and in the mid-time regime $0 < r,l \ll a \ll t \ll b$, we arrive at the following result:
\begin{align}\label{mixedlatetime}
S_A^\text{mixed} =\frac{c}{6}\log\left(\frac{\beta b^2 t}{2\pi\epsilon^2s}\sinh\left(\frac{2\pi^2}{\beta}\right)\right).
\end{align}
Note that this logarithmic growth of entanglement entropy $\Delta S_A\sim \frac{c}{6}\log t$ is also seen for local operator quenches (\ref{logwlq}). This is because the mixed-state excitation is a canonical distribution of local operator excitations as worked out in \cite{Bhattacharyya:2019ifi}.

Similar computation can be done for the low-temperature phase to obtain
\begin{equation}
    S_A^\text{mixed} = S_A^\text{vac} = \frac{c}{3}\log\left(\frac{b-a}{\epsilon}\right).
\end{equation}

\subsection{Duality of single local quenches}\label{dualitysingle}

Comparing (\ref{logwlq}) from the previous section, where we reviewed single local operator quenches, and (\ref{mixedlatetime}), we can notice similarities beyond their logarithmic time dependence. Replacing $2\pi/\beta$ by $i\alpha_H$ and $s$ by $\delta$ in (\ref{mixedlatetime}), we obtain
\begin{equation}
    S_A^\text{pure} =\frac{c}{6}\log\left(\frac{b^2 t}{\alpha_H\epsilon^2\delta}\sin\left(\alpha_H\pi\right)\right),
\end{equation}
which is (\ref{logwlq}) plus the entanglement entropy in the vacuum. This suggests an equivalence between local operator (pure-state) quenches and mixed-state quenches.

The expressions (\ref{logwlq}) and (\ref{mixedlatetime}) only hold in the limit of a large (semi-infinite) subsystem and at late times, but it turns out we can demonstrate the equivalence beyond those limits. Substituting (\ref{zevo}) and (\ref{zbevo}) into the general expression (\ref{HHLLformula}) of $\Delta S_A$ for a single local operator quench and adding
\begin{equation}
    S_A^\text{vac} = \frac{c}{3}\log\left(\frac{b-a}{\epsilon}\right)
\end{equation}
yields the same result as substituting (\ref{torustransf}) and (\ref{torustransfb}) into the general expression (\ref{BTZ_EE}) of $S_A^\text{mixed}$, but again with the necessary replacements
\begin{align}
    i\alpha_H \leftrightarrow \frac{2\pi}{\beta}, \qquad \delta \leftrightarrow s. \label{idenfyl}
\end{align}

Given this correspondence between the entanglement entropy of these two different classes of local quench, one might wonder whether we have a more general equivalence between operators:
\ba
\Psi^\text{pure}_{{\alpha_H},\delta}(x,t)\simeq 
\Psi^\text{mixed}_{\beta,s}(x,t) \label{dualityr}
\ea
via the identification (\ref{idenfyl}). $\Psi^\text{pure}_{{\alpha_H},\delta}(x,t)$ is the operator that creates the bra and ket states of a local operator excitation, namely
\ba
\Psi^\text{pure}_{{\alpha_H},\delta}(x,t)
=\mathcal{O}^{\dagger}(x,\delta-it)\mathcal{O}(x,-\delta-it).
\ea
On the other hand, $\Psi^\text{mixed}_{\beta,s}(x,t)$ is the operator that creates a mixed-state local excitation  (\ref{mixedexcitation}) from the vacuum. It should be noted that (\ref{dualityr}) is to be understood to hold within correlation functions. For example, in the presence of other local operators $\mathcal{Q}_i$, the equivalence would imply
\ba
&& \la \Psi^\text{pure}_{{\alpha_{H1}},\delta_1}(x_1,t_1)
\Psi^\text{pure}_{{\alpha_{H2}},\delta_2}(x_2,t_2)\cdots
\mathcal{Q}_a(x_a,t_a)\mathcal{Q}_b(x_b,t_b)\cdots\lb \no
&& =\la \Psi^\text{mixed}_{\beta_1,s_1}(x_1,t_1)
\Psi^\text{mixed}_{{\beta_{2}},s_2}(x_2,t_2)\cdots
\mathcal{Q}_a(x_a,t_a)\mathcal{Q}_b(x_b,t_b)\cdots\lb,
\label{dualityrcf}
\ea
where $i\ap_{Hi}=2\pi/\beta_i$ and $\delta_i=s_i$. If this general equivalence holds, we essentially have a correspondence between pure-state and mixed-state black holes, as local quenches realized by an operator with conformal weight $h_\mathcal{O}>c/24$ are dual to black hole microstates \cite{Nozaki:2013wia,Asplund:2014coa}. We can also see this as a consequence of the eigenstate thermalization hypothesis (ETH) \cite{Srednicki:1994mfb}, generalizing the argument that was made for pure-state AdS black holes in \cite{Fitzpatrick:2014vua}.

Furthermore, the correspondence would imply that the system currently under consideration, consisting of a local operator quench and a mixed-state quench, can be recast as one with the locations of quenches swapped. This idea will be tested towards the end of this section, as we can compare analytic results obtained in the limits $\delta \ll s$ and $s \ll \delta$. A perhaps more ambitious and not-so-easy-to-check claim is that a system consisting of $n$ local operator quenches, whose entanglement entropy is expressed in terms of a $(2n+2)$-point correlation function, may be equivalent to one consisting of $n$ mixed-stated quenches, whose entanglement entropy is expressed in terms of a two-point function on a genus-$n$ surface. However, it must be noted that the equivalence (\ref{dualityrcf}) is expected to hold only when $\mathcal{Q}_i$ are strictly local operators, as a coarse-grained observer cannot distinguish between pure-state and mixed-state excitations. On the other hand, if $\mathcal{Q}_i$ are twist operators of the replicated CFT, this equivalence may only hold `formally' because the entanglement entropy is a fine-grained quantity that can distinguish whether a given state is pure or mixed. For example, if we consider the early-time behavior of entanglement entropy for a case II setup, we see that $S_A^\text{pure} = S_A^\text{vac}$ ($\Delta S_A = 0$) for local operator quenches, but $S_A^\text{mixed} = S_A^\text{vac} + S_A^\text{thermal}$ for mixed-state quenches. This mismatch occurs because different channels are considered for local operator quenches, whereas $(\Theta,\bar{\Theta})$ is uniquely determined for mixed-state quenches. In fact, we can check that $S_A^\text{pure} = S_A^\text{vac} + \Delta S_{A,1}$, obtained solely from channel 1, is precisely dual to $S_A^\text{mixed}$ at all times. This formal equivalence (or lack thereof) can be understood in the holographic picture as being caused by the different homology constraints that are imposed when dealing with pure/mixed states.

\subsection{Insertion of local operator excitation}\label{4ptfunc}

In the previous subsection, we found that mixed-state excitations correspond to introducing an identified set of finite-sized holes in the Euclidean geometry of the theory. Another important takeaway from the section was the special property of holographic theories that allows us to treat this geometry, which naively is conformally equivalent to a torus, as instead being equivalent to an infinitely long cylinder. We can effectively decompactify one of the directions of spacetime, giving rise to two distinct phases --- BTZ black hole and thermal AdS --- depending on our choice of direction. What we would like to do now is to consider local operator excitations on this geometry. On a technical level, this calculation is similar to the one carried out in \cite{Caputa:2014eta}, where the local operator quench was studied at finite temperature. In this paper, we shall mainly be interested in the BTZ phase, where the spatial direction is decompactified, as it gives physically nontrivial results. The thermal AdS phase is discussed in appendix \ref{ap:thermal}, where we shall see that results are similar to what we have already seen for single local operator excitations.

At this point, it is natural to question whether the cylindrical approximation of the geometry holds up in the presence of heavy operators. In section \ref{validity}, we will argue that the approximation is valid when the regulators $\delta,s$ of quenches are not of the same order, i.e. $\delta/s \ll 1$ or $\delta/s \gg 1$. We shall therefore limit our discussion in this section to those two particular limits of regulators. Taking these limits also allows us to compute simple analytical expressions for entanglement entropy --- something we always like.

\begin{figure}
    \centering
    \includegraphics[width=0.7\linewidth]{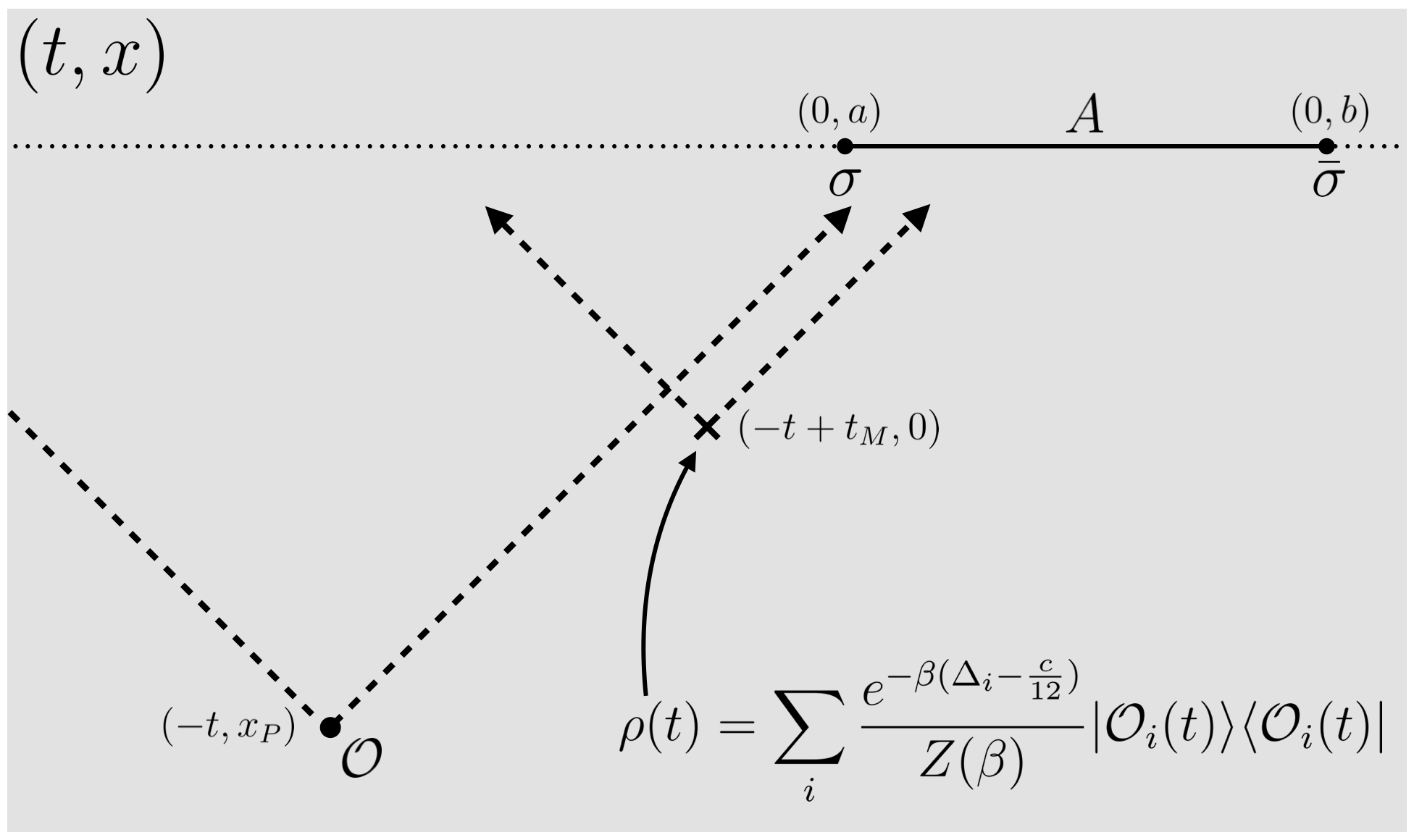}
    \caption{Our setup of interest depicted on the Lorentzian plane. The local operator excitation by ${\cal O}$ produces an entangled pair of modes that propagate in opposite directions at the speed of light. When $t_M+x_P=0$, one of the modes directly collides with the mixed-state excitation.}
    \label{fig:LorentzianSetup}
\end{figure}

In addition to the mixed-state excitation at $(x,t)=(0,t_M)$ with $t_M>0$, consider adding a local operator excitation $\mathcal{O}$ at $(x,t)=(x_P,0)$, as depicted in figure \ref{fig:LorentzianSetup}. When $x_P\pm t_M=0$, one of the entangled pair of modes produced by the local operator excitation collides with the mixed-state excitation, where we expect an interesting interplay. Without loss of generality, assume $x_P\le 0$ hereafter (so lightlike separation is only attainable when $x_P+t_M=0$). Once we pass this on to the Euclidean plane, the locations of pure-state and mixed-state excitations are as shown in figure \ref{fig:setup}.

Calculation of entanglement entropy requires the computation of four-point function $\langle \mathcal{O}^\dagger \mathcal{O} \sigma_n\bar{\sigma}_n \rangle$, which can be done via the HHLL approximation as previously seen in section \ref{HHLL}. Our strategy is therefore straightforward: we must first conformally transform the two-hole geometry in figure \ref{fig:setup} into an annulus where the time direction is compactified and the space direction extends radially. This was done partially in the previous subsection, where we made use of the conformal transformation that mapped the two-hole geometry onto a finite strip. A further application of the exponential map with the correct power should yield the desired annulus. With the space direction extending radially, we can now reap the benefits of working with a holographic theory and ignore the two boundaries of the annulus, which means that the geometry is now the full Euclidean plane. At this point, we can simply apply the method of HHLL approximation and compute the four-point function by tracking the behavior of cross ratios of insertion points.

\begin{figure}[h]
    \centering
    \includegraphics[width=0.7\linewidth]{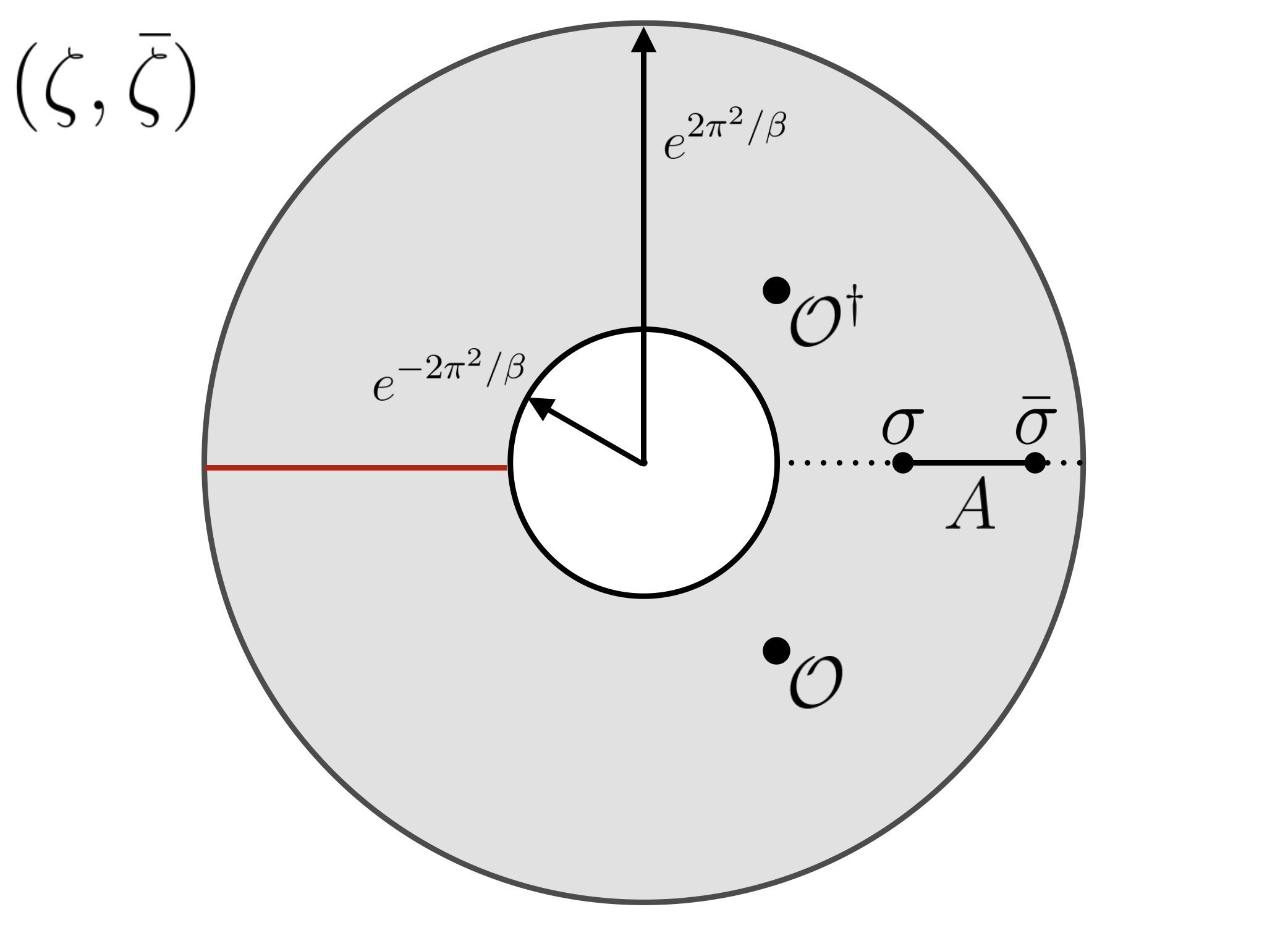}
    \caption{Conformally transforming our initial two-hole geometry would eventually give us a torus (drawn as an annulus with its boundaries compactified), where the time direction is compactified tangentially and the space direction is compactified radially. The two identified holes in figure \ref{fig:setup} are mapped onto the red segment.}
    \label{fig:maptoannulus}
\end{figure}

We have already seen in (\ref{torustransf}) and (\ref{torustransfb}) that the map
\begin{align}
    w &= \log\left(-\frac{X-t+t_M+is}{X-t+t_M-is}\right), \\
    \bar{w} &= \log\left(-\frac{\bar{X}+t-t_M-is}{\bar{X}+t-t_M+is}\right),
\end{align}
maps our two-hole geometry to a finite-sized strip. Its real direction corresponds to time with periodicity $\beta$ and its imaginary direction corresponds to space with periodicity $2\pi i$. In order to compactify time in the tangential direction of an annulus, we must apply the following exponential map:
\begin{align}
    \zeta &= e^{\frac{2\pi i}{\beta}w} \nonumber\\
    &= \left(-\frac{X-t+t_M+is}{X-t+t_M-is}\right)^\frac{2\pi i}{\beta}, \\
    \bar{\zeta} &= e^{-\frac{2\pi i}{\beta}\bar{w}} \nonumber\\
    &= \left(-\frac{\bar{X}+t-t_M-is}{\bar{X}+t-t_M+is}\right)^{-\frac{2\pi i}{\beta}}.\label{zetamapp}
\end{align}
This brings us to the geometry depicted in figure \ref{fig:maptoannulus}. Our spatial subsystem $A$ now extends radially, as expected. For convenience, let us define the following variables, which denote the location of operator insertions in our new geometry:
\begin{align}
    \zeta_{\sigma_n} &\coloneq \zeta(a), \\
    \zeta_{\bar{\sigma}_n} &\coloneq \zeta(b), \\
    \zeta_{\mathcal{O}} &\coloneq \zeta(-i(\delta+it)+x_P), \\
    \zeta_{\mathcal{O}^\dagger} &\coloneq \zeta(i(\delta-it)+x_P).
\end{align}
The barred variables $\bar{\zeta}_{\sigma_n},\bar{\zeta}_{\bar{\sigma}_n},\bar{\zeta}_{\mathcal{O}},\bar{\zeta}_{\mathcal{O}^\dagger}$ are also defined in a similar manner.

As explained earlier, in the BTZ phase of a holographic theory, this torus is effectively a decompactified annulus that extends to a full Euclidean plane with no boundaries at $|\zeta|=e^{\pm 2\pi^2/\beta}$. On this plane, we shall apply the method of HHLL approximation, which requires us to compute the cross ratios $(z,\bar{z})$ of insertion points. Here, we will take the limit $b \to \infty$ because we are ultimately interested in taking the subsystem to be a half line (and also because we want to simplify our calculations).

The full expression for $z$ is given by
\begin{align}\label{zanalytic}
    z &= \frac{(\zeta_{\sigma_n}-\zeta_{\bar{\sigma}_n})(\zeta_{\mathcal{O}^\dagger}-\zeta_{\mathcal{O}})}{(\zeta_{\sigma_n}-\zeta_{\mathcal{O}^\dagger})(\zeta_{\bar{\sigma}_n}-\zeta_{\mathcal{O}})} \nonumber\\
    &= \frac{\left(\left(-\frac{a-t+t_M+is}{a-t+t_M-is}\right)^{\frac{2\pi i}{\beta}}-\left(-\frac{b-t+t_M+is}{b-t+t_M-is}\right)^{\frac{2\pi i}{\beta}}\right)}{\left(\left(-\frac{a-t+t_M+is}{a-t+t_M-is}\right)^{\frac{2\pi i}{\beta}}-\left(-\frac{x_P+t_M+i\delta+is}{x_P+t_M+i\delta-is}\right)^{\frac{2\pi i}{\beta}}\right)} \nonumber\\
    &\qquad\qquad\qquad\qquad\qquad
    \cdot\frac{\left(\left(-\frac{x_P+t_M+i\delta+is}{x_P+t_M+i\delta-is}\right)^{\frac{2\pi i}{\beta}}-\left(-\frac{x_P+t_M-i\delta+is}{x_P+t_M-i\delta-is}\right)^{\frac{2\pi i}{\beta}}\right)}{\left(\left(-\frac{b-t+t_M+is}{b-t+t_M-is}\right)^{\frac{2\pi i}{\beta}}-\left(-\frac{x_P+t_M-i\delta+is}{x_P+t_M-i\delta-is}\right)^{\frac{2\pi i}{\beta}}\right)}.
\end{align}
Likewise the full expression for $\bar{z}$ is given by
\begin{multline}\label{zbaranalytic}
    \bar{z} = \frac{\left(\left(-\frac{-a-t+t_M+is}{-a-t+t_M-is}\right)^{-\frac{2\pi i}{\beta}}-\left(-\frac{-b-t+t_M+is}{-b-t+t_M-is}\right)^{-\frac{2\pi i}{\beta}}\right)}{\left(\left(-\frac{-a-t+t_M+is}{-a-t+t_M-is}\right)^{-\frac{2\pi i}{\beta}}-\left(-\frac{-x_P+t_M+i\delta+is}{-x_P+t_M+i\delta-is}\right)^{-\frac{2\pi i}{\beta}}\right)} \\
    \cdot\frac{\left(\left(-\frac{-x_P+t_M+i\delta+is}{-x_P+t_M+i\delta-is}\right)^{-\frac{2\pi i}{\beta}}-\left(-\frac{-x_P+t_M-i\delta+is}{-x_P+t_M-i\delta-is}\right)^{-\frac{2\pi i}{\beta}}\right)}{\left(\left(-\frac{-b-t+t_M+is}{-b-t+t_M-is}\right)^{-\frac{2\pi i}{\beta}}-\left(-\frac{-x_P+t_M-i\delta+is}{-x_P+t_M-i\delta-is}\right)^{-\frac{2\pi i}{\beta}}\right)}.
\end{multline}
Notice that we have two different UV cutoffs $\delta, s$, corresponding to cutoffs for the local operator and mixed-state quench respectively. Depending on how we take the limits $\delta, s \to 0$, factors such as $\zeta_{\mathcal{O}}$ and $\zeta_{\mathcal{O}^\dagger}$ will assume different values. We shall treat both cases of $\delta<s$ and $s<\delta$, the former being more straightforward than the latter. The apparent issue with the $s<\delta$ regime is that $\zeta_{\mathcal{O}}, \zeta_{\mathcal{O}^\dagger} \to (-1)^{\frac{2\pi i}\beta}$ as $s \ll \delta$, but there is no way of deciding their logarithmic branches. We will return to this issue in section \ref{mixedgeneral}, where we discuss this ambiguity in detail and see how different choices affect the computation of $\Delta S_A$.

Finally, the application of (\ref{HHLLformula}) gives us $\Delta S_A$. However, it must be stressed that $\Delta S_A$ is only the increase in entanglement entropy as measured from the `vacuum', which, in this setup, incorporates the mixed-state local quench and its contribution to entanglement entropy. The full computation of $S_A$ requires one to combine the contribution from (\ref{BTZ_EE}) with $\Delta S_A$, yielding
\begin{align}
    S_A &= S_A^\text{mixed} + \Delta S_A \nonumber\\
    &= \frac{c}{6}\log\left(\frac{\beta^2}{\pi^2 \epsilon_a \epsilon_b}\sinh\left(\frac{\pi(\Theta_b-\Theta_a)}{\beta}\right)\sinh\left(\frac{\pi(\bar{\Theta}_b-\bar{\Theta}_a)}{\beta}\right)\right) \nonumber\\
    & \qquad\qquad+ \frac{c}{6}\log\left(\frac{\sin\left(\frac{\alpha_H \theta}{2}\right)\sin\left(\frac{\alpha_H \bar{\theta}}{2}\right)}{\alpha^2_H \sin\left(\frac{\theta}{2}\right)\sin\left(\frac{\bar{\theta}}{2}\right)}\right).
\end{align}

\subsection{When $\delta \ll s$}\label{purelimit}
Let us first take the separation between excitations to be lightlike, i.e. $x_P+t_M=0$, where one of the entangled pair of modes created by the local operator directly collides with the mixed-state excitation. This not only simplifies our analytic computation, but we also anticipate the interaction between the quenches to be maximal at this separation. We also shall assume a case I setup (as defined in section \ref{singleinsertion}), where $a>x_P$.

\begin{figure}[h]
    \centering
    \includegraphics[width=0.45\linewidth]{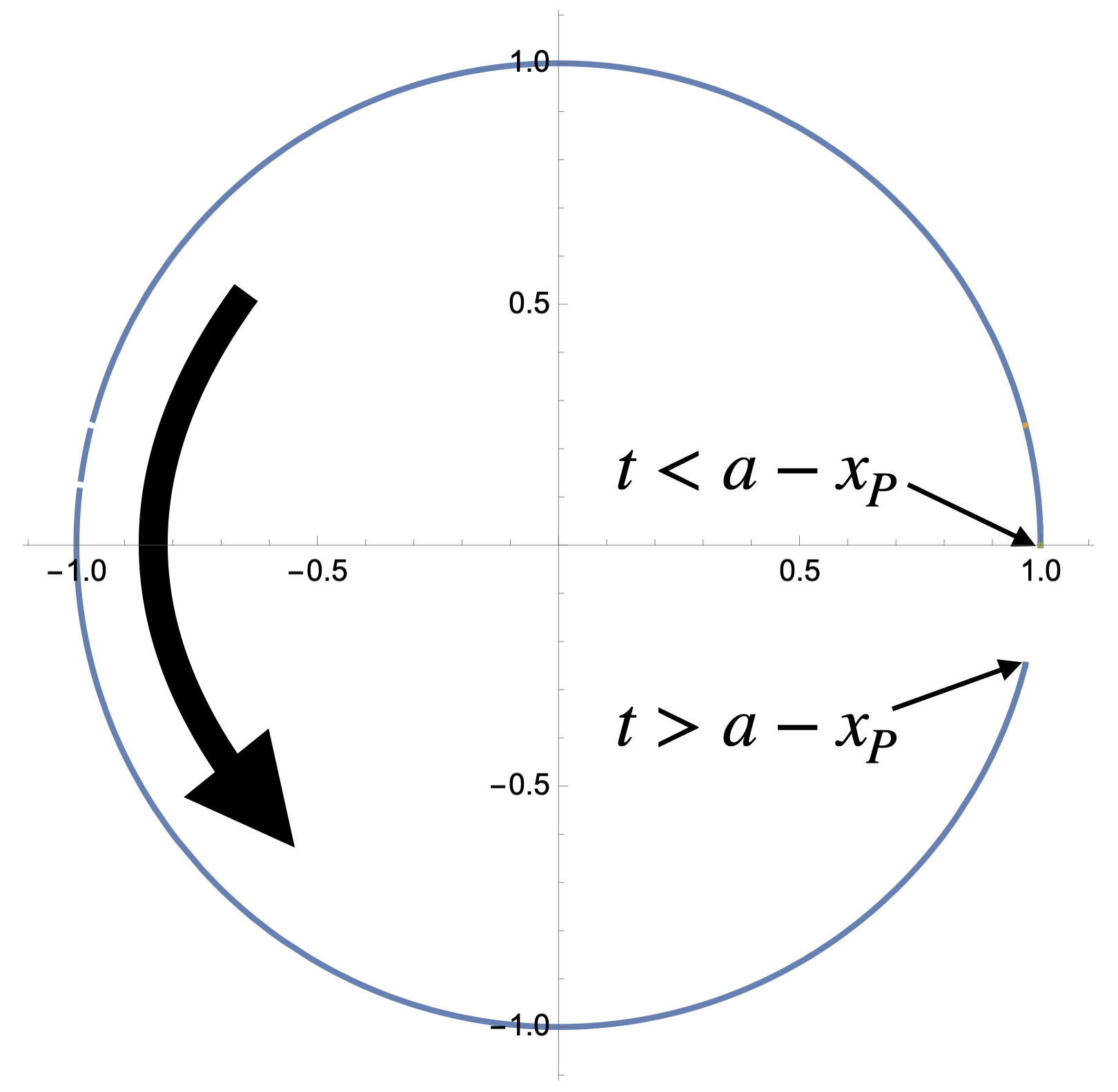}
    \caption{Behavior of $1-z$ as time evolves. Starting from $1-z=1$, it works its way around the origin anticlockwise back to $1-z=1$ (the circle should close once the regulators are taken to 0). In the large-$t$ limit, $1-z$ should therefore be replaced by $e^{2\pi i}(1-z)$, with an additional factor that accounts for its monodromy.}
    \label{fig:z_behavior}
\end{figure}

\begin{figure}[h]
    \centering
    \includegraphics[width=0.5\linewidth]{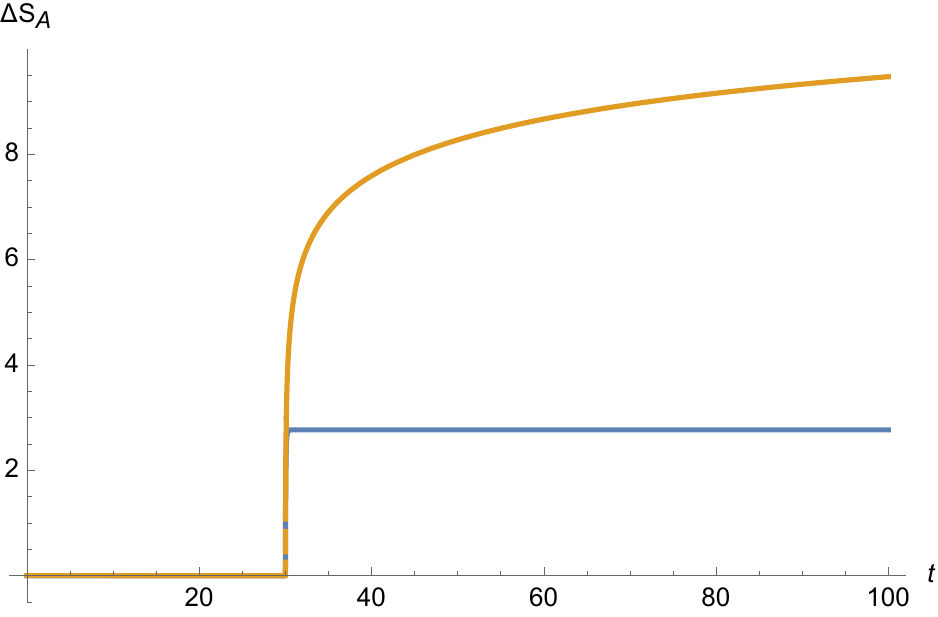}
    \caption{Increase in entanglement entropy due to a single pure-state local operator quench. Subsystem is taken to be $a=20, b=1000$, and the insertions are at $x_P=-10$ and $t_M=10$. The yellow line indicates a logarithmic time evolution that is observed when the quench takes place in a vacuum background. In the presence of another quench, notably a mixed-state local operator quench, the contribution is heavily suppressed and is reduced to a constant bump, as indicated by the blue line. The regulators are taken to be $\delta=0.01,s=1$. $\beta=1$ (BTZ phase). }
    \label{fig:entsup}
\end{figure}

In the limit $\delta \ll s$ and $b \to \inf$, we obtain
\begin{equation}\label{BTZ_z}
    z = \frac{\left(\left(-\frac{a-t+t_M+is}{a-t+t_M-is}\right)^{\frac{2\pi i}{\beta}}-e^\frac{2\pi^2}{\beta}\right)\left(\frac{8\pi i\delta}{\beta s}\right)}{\left(\left(-\frac{a-t+t_M+is}{a-t+t_M-is}\right)^{\frac{2\pi i}{\beta}}-1\right)\left(e^\frac{2\pi^2}{\beta}-1\right)}.
\end{equation}
Let us first observe the large-$t$ behavior of (\ref{BTZ_z}), then look at the time evolution $1-z$ to find its monodromy. When $a \ll t$, $-\frac{a-t+t_M+is}{a-t+t_M-is}$ approaches $e^{\pi i}$, so
\begin{align}    
    z &= \frac{\left(e^{-\frac{2\pi^2}{\beta}}-e^\frac{2\pi^2}{\beta}\right)\left(\frac{8\pi i\delta}{\beta s}\right)}{\left(e^{-\frac{2\pi^2}{\beta}}-1\right)\left(e^\frac{2\pi^2}{\beta}-1\right)} \nonumber \\
    &= i\frac{8\pi\delta}{\beta s \tanh\left(\frac{\pi^2}{\beta}\right)} + O(t^{-1}),
\end{align}
or equivalently,
\begin{equation}
    \theta = 2\pi m - \frac{8\pi\delta}{\beta s \tanh\left(\frac{\pi^2}{\beta}\right)} + O(t^{-1}),
\end{equation}
where $m=0$ or $1$, depending on its monodromic behavior ($m=-1$ is not possible as the permissible range of $\theta$ is $-2\pi < \theta < 2\pi$). It turns out, as can be seen in figure \ref{fig:z_behavior}, $1-z$ revolves around the origin once in the anticlockwise direction as it evolves with time. Therefore $1-z$ in the large-$t$ limit does not simply approach 1, but $e^{2\pi i}$ instead; the correct choice of $n$ would be $m=1$.
The large-$t$ behavior of $\bar{z}$ comes out as
\begin{align}
    \bar{z} &= \frac{\left(\left(-\frac{-a-t+t_M+is}{-a-t+t_M-is}\right)^{-\frac{2\pi i}{\beta}}-e^\frac{2\pi^2}{\beta}\right)\left(-e^{-\frac{2\pi^2}{\beta}}\frac{2\pi i\delta s}{\beta t_M^2}\right)}{\left(\left(-\frac{-a-t+t_M+is}{-a-t+t_M-is}\right)^{-\frac{2\pi i}{\beta}}-e^{-\frac{2\pi^2}{\beta}}\right)\left(e^\frac{2\pi^2}{\beta}-e^{-\frac{2\pi^2}{\beta}}\right)} \nonumber\\
    &= \frac{\left(-e^\frac{2\pi^2}{\beta}\frac{4\pi s}{\beta t}\right)\left(-e^{-\frac{2\pi^2}{\beta}}\frac{2\pi i\delta s}{\beta t_M^2}\right)}{\left(e^\frac{2\pi^2}{\beta}-e^{-\frac{2\pi^2}{\beta}}\right)\left(e^\frac{2\pi^2}{\beta}-e^{-\frac{2\pi^2}{\beta}}\right)} \nonumber\\
    &= i\frac{2\pi^2\delta s^2}{\beta^2 t_M^2 \sinh^2\left(\frac{2\pi^2}{\beta}\right)t} + O(t^{-2}).
\end{align}
It can be checked that $1-\bar{z}$ does not exhibit any monodromy under time evolution, so
\begin{equation}
    \bar{\theta} = \frac{2\pi^2\delta s^2}{\beta^2 t_M^2 \sinh^2\left(\frac{2\pi^2}{\beta}\right)t} + O(t^{-2}).
\end{equation}
Finally, making use of (\ref{HHLLformula}) or, equivalently, (\ref{HHLLsimple}), the change in entanglement entropy due to the insertion of local operators can be computed as
\begin{equation}\label{EEpure}
    \Delta S_A = \frac{c}{6} \log \left(\frac{\beta s}{4\pi \alpha_H \delta}\sin(\alpha_H \pi)\tanh\left(\frac{\pi^2}{\beta}\right)\right).
\end{equation}
This is a peculiar result, given that at leading order $\Delta S_A$ displays no time dependence at all, as can be clearly seen in figure \ref{fig:entsup}. Since there is a suppression of what used to be a logarithmic growth of entanglement entropy, we shall call this phenomenon `entanglement suppression'.

It is possible to provide a physical explanation as to why we observe entanglement suppression. In this setup of lightlike-separated local quenches, we expect one of the entangled pair of modes created by the local operator to be absorbed by the mixed-state excitation and converted into an excitation in another Hilbert space, which is obtained by purifying the mixed-state excitation just like in the case of a thermofield double. This can be ascribed to strong interactions that are typical in holographic CFTs. Therefore, this phenomenon is directly related to the chaotic dynamics of holographic CFTs, which, in their dual picture, can be thought of as scattering of gravitational waves off heavy objects.

From (\ref{EEpure}), we also find that $\Delta S_A$ becomes large when $\delta$ is small and this is explained by the fact that the energy of local operator excitation scales as $1/\delta$. Moreover, $\Delta S_A$ decreases as $s$ gets smaller. This also agrees with the expectation that the entanglement suppression should be more enhanced for smaller values of $s$.

Let us now take the separation between excitations to be either timelike or spacelike so that $|x_P+t_M| \gg \delta,s$. When $x_P+t_M>0$,
\begin{align}    
    z &= \frac{\left(e^{-\frac{2\pi^2}{\beta}}-e^\frac{2\pi^2}{\beta}\right)\left(e^{\frac{2\pi^2}{\beta}}\frac{8\pi i\delta s}{\beta (x_P+t_M)^2}\right)}{\left(e^{-\frac{2\pi^2}{\beta}}-e^{\frac{2\pi^2}{\beta}}\right)\left(e^{\frac{2\pi^2}{\beta}}\frac{4\pi s}{\beta (x_P+t_M)}\right)} \nonumber \\
    &= i\frac{2\delta}{x_P+t_M} + O(t^{-1}),
\end{align}
and when $x_P+t_M<0$,
\begin{align}    
    z &= \frac{\left(e^{-\frac{2\pi^2}{\beta}}-e^\frac{2\pi^2}{\beta}\right)\left(e^{-\frac{2\pi^2}{\beta}}\frac{8\pi i\delta s}{\beta (x_P+t_M)^2}\right)}{\left(e^{-\frac{2\pi^2}{\beta}}\frac{4\pi s}{\beta (x_P+t_M)}\right)\left(e^\frac{2\pi^2}{\beta}-e^{-\frac{2\pi^2}{\beta}}\right)} \nonumber \\
    &= -i\frac{2\delta}{x_P+t_M} + O(t^{-1}).
\end{align}
Similarly, $\bar{z}$ comes out as
\begin{equation}
    \bar{z}=i\frac{8\pi^2\delta s^2}{\beta^2 (-x_P+t_M)^2 \sinh^2\left(\frac{2\pi^2}{\beta}\right)t} + O(t^{-2}).
\end{equation}
Therefore,
\begin{equation}\label{nonlightlike}
    \Delta S_A = \frac{c}{6}\log\left(\frac{|x_P+t_M|}{\alpha_H\delta}\sin(\alpha_H \pi)\right).
\end{equation}

\begin{figure}[h]
  \centering
    \includegraphics[width=0.5\linewidth]{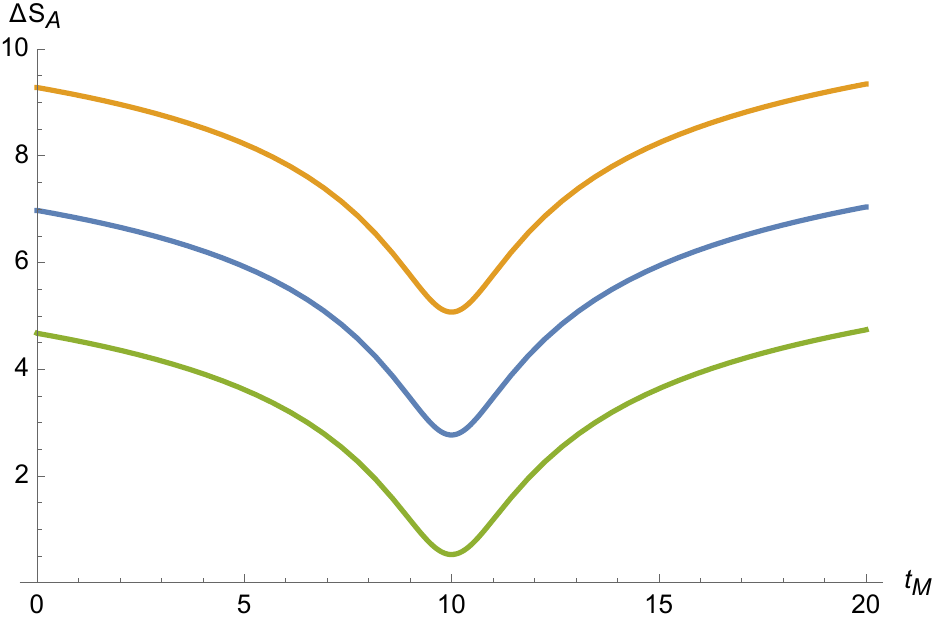}
  \caption{Plots of $\Delta S_A$ as a function of $t_M$, given $x_P = -10$. We find that $\Delta S_A$ is minimized at $t_M+x_P=0$. Subsystem $A$ is taken to be sufficiently large, $a=20, b=1000$, and time is taken to be $t=100$. The regulators are taken to be $\delta=0.001,s=1$ in the orange plot, $\delta=0.01,s=1$ in the blue plot, and $\delta=0.1,s=1$ in the green plot. $\beta=1$ (BTZ phase) for all plots.} 
\label{fig:caseIImixedfixed}
\end{figure}

First of all, readers should be reminded that when $|x_P+t_M|$ is very small, the entanglement entropy is correctly described by the previous result (\ref{EEpure}), so they need not fret about the divergence in (\ref{nonlightlike}) at $|x_P+t_M|=0$. More importantly, just like the case where the separation was taken to be lightlike, $\Delta S_A$ has no time-dependence at leading order, so entanglement suppression occurs regardless of how the excitations are located relative to each other. We also find that $\Delta S_A$ grows as $|x_P+t_M|$ moves away from $0$. Indeed, as plotted in figure \ref{fig:caseIImixedfixed}, the suppression effect is maximized when the two excitations are lightlike-separated, i.e. $x_P+t_M=0$, as one of the entangled pair of modes collides with the mixed-state excitation.

\begin{figure}[ttt]
  \centering
   \includegraphics[width=0.7\linewidth]{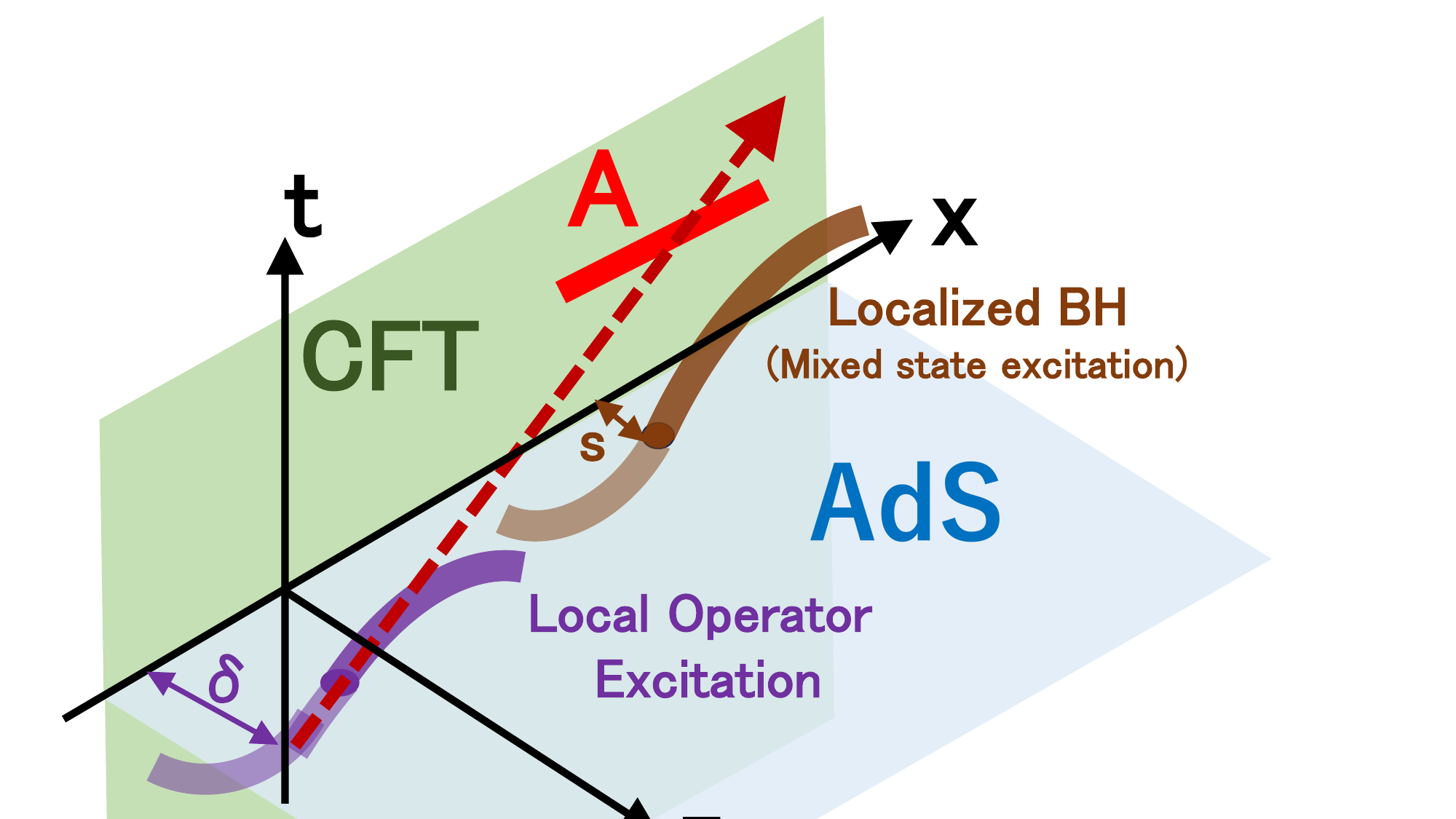}
  \caption{A sketch of the gravitational dual of our setup with a mixed-state and a local operator excitation. The green-colored plane describes the AdS boundary where the CFT lives and the blue plane is the time slice $t=0$ of the bulk AdS$_3$. The purple thick curve describes the core of a deficit angle dual to the local operator excitation, while the brown curve describes a localized black hole dual to the mixed-state excitation. The red arrows show the propagation of gravitational waves emitted by the local operator excitation, some of which should reach subsystem $A$.}
\label{fig:holsketch}
\end{figure}

As usual in the AdS/CFT, this holographic CFT calculation is equivalently described by the gravitational dual of AdS$_3$. As sketched in figure \ref{fig:holsketch}, the local operator and the mixed-state excitation are dual to a deficit angle and a localized black hole in AdS$_3$. The analysis of holographic local operator quenches \cite{Nozaki:2013wia,Bhattacharyya:2019ifi} shows that their cores are located at $z=\s{\delta^2+t^2}$ and $z=\s{s^2+(t-t_M)^2}$ respectively under time evolution, working with the Poincar\'e AdS$_3$ metric $ds^2=z^{-2}(dz^2-dt^2+dx^2)$. The propagation of the entangled pair created by the local operator is dual to that of the gravitational shockwave produced by the time-dependent deficit angle. We expect that some of this shockwave is absorbed by the localized black hole, dual to the mixed-state quench. This provides a gravitational interpretation of entanglement suppression.

\begin{figure}[hhh]
    \centering
    \includegraphics[width=0.32\linewidth]{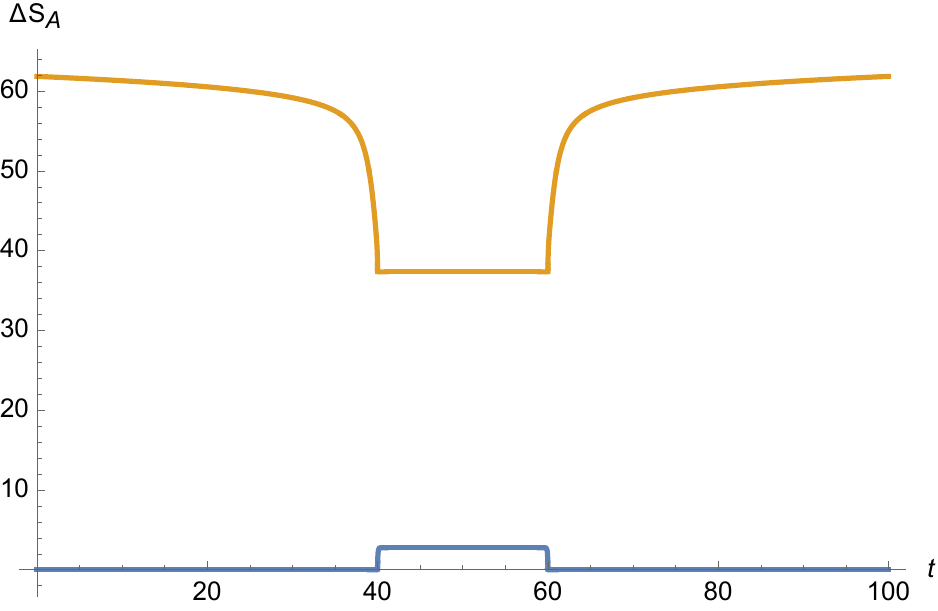}
    \includegraphics[width=0.32\linewidth]{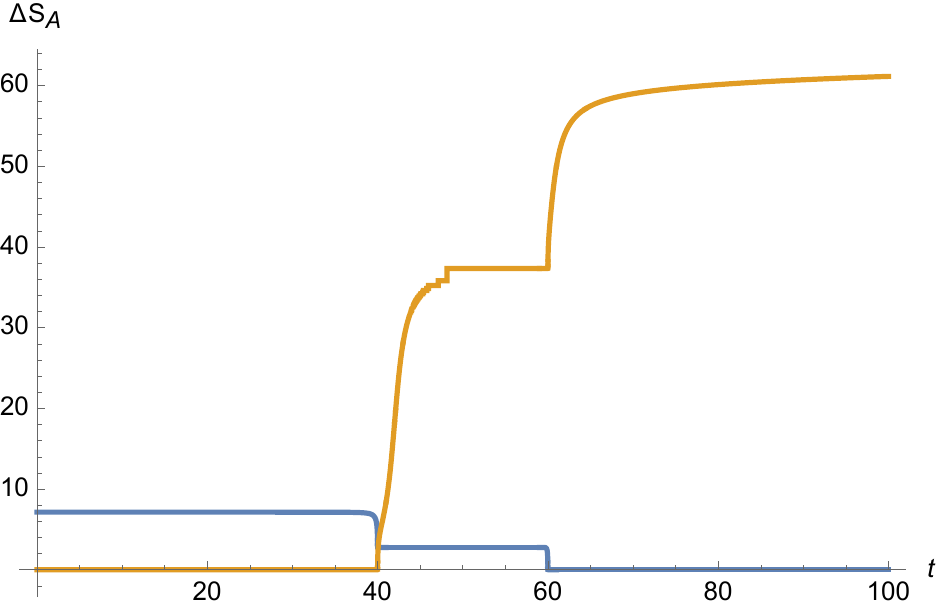}
    \includegraphics[width=0.32\linewidth]{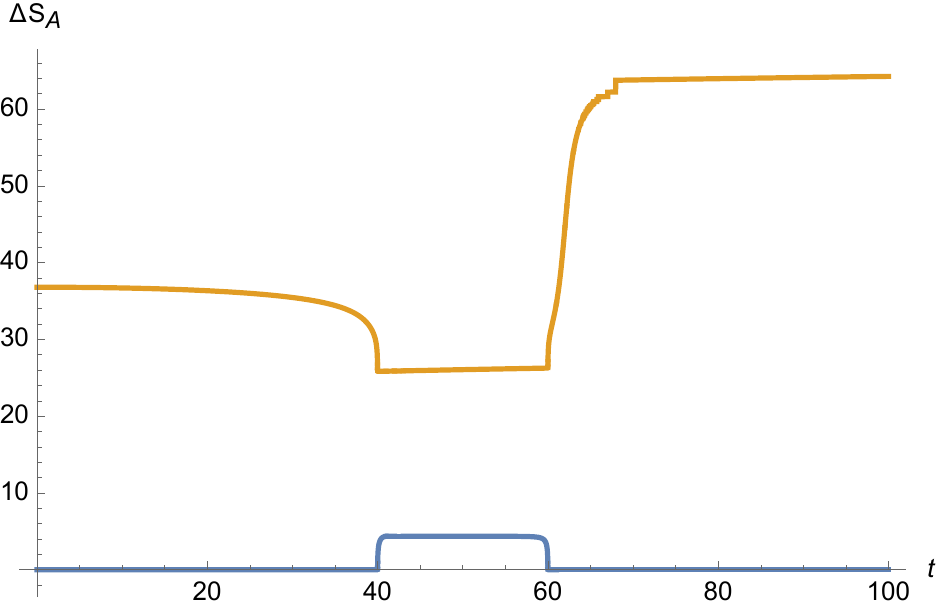}
    \caption{Contributions to entanglement entropy from different channels, depending on the position of the pure-state excitation relative to the subsystem. For all graphs, we have set $x_P=-1$ and $t_M=1$, so the separation between excitations are lightlike. In the left graph, $a=39,b=59$, so the excitation is placed to the left of the subsystem. In the middle graph, $a=-41,b=59$, so the excitation is placed inside the subsystem. In the right graph, $a=-61,b=-41$, so the excitation is placed to the right of the subsystem. They all result in the same time evolution of entanglement entropy, but channels contribute differently in each of the three cases. The regulators are taken to be $\delta=0.01,s=1$. $\beta=1$ (BTZ phase).}
    \label{fig:cases}
\end{figure}

\begin{figure}[hhh]
  \centering
  \includegraphics[width=0.4\linewidth]{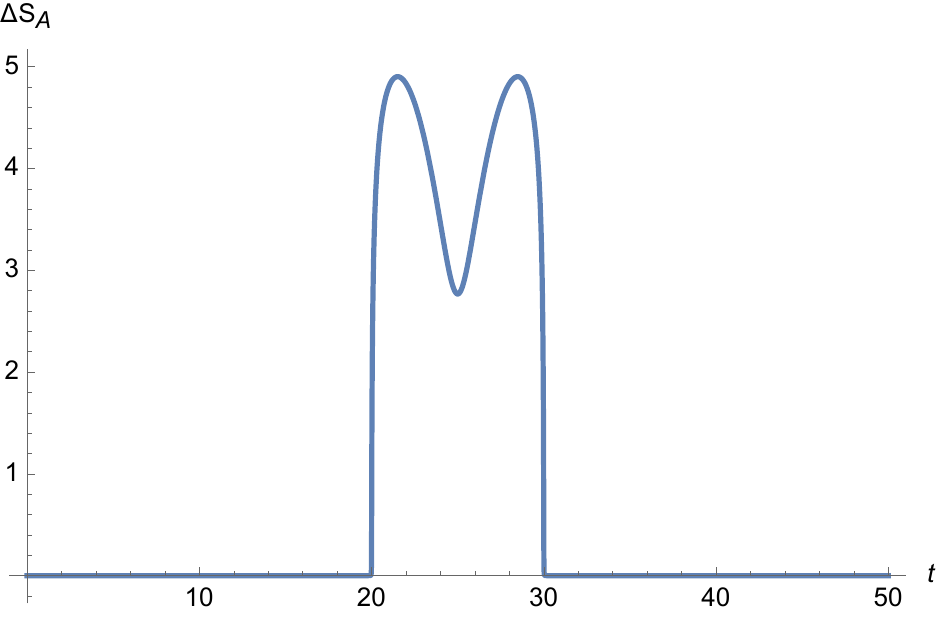}
   \includegraphics[width=0.4\linewidth]{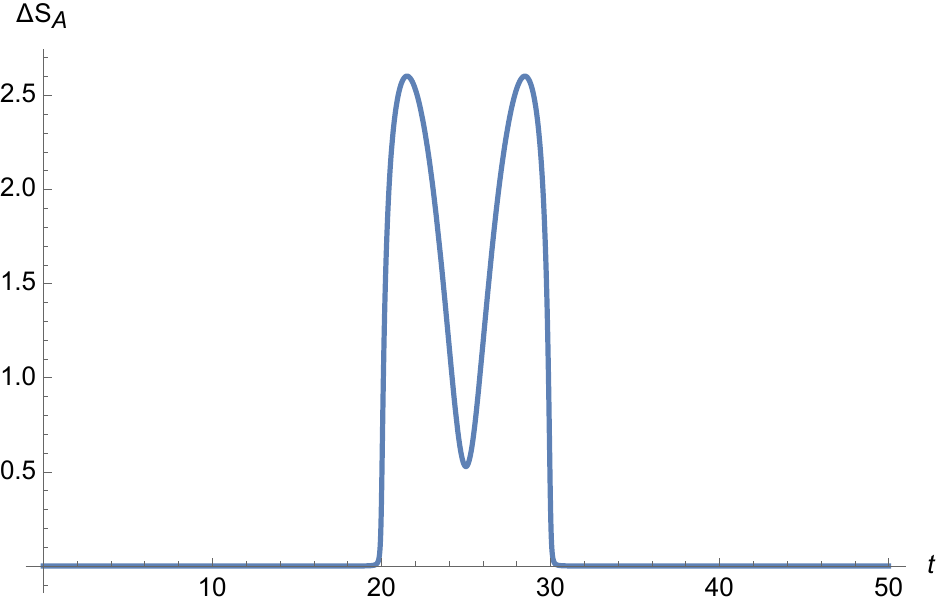}
  \caption{The plots $\Delta S_A$ as a function of $t$ 
  for $t_M=t-15$ and $x_P=-10$ for $\delta=0.01$ (left panel) and $\delta=0.1$ (right panel). We chose the subsystem to be $A=[10,20]$. The entanglement suppression is maximized at $t=25$. The mixed-state regulator is fixed at $s=1$. $\beta=1$ (BTZ phase).}
\label{fig:MLQ1}
\end{figure}

We can move away from the limit of the subsystem being semi-infinite and consider the time evolution of entanglement entropy for finite subregions. As touched upon in section \ref{singleinsertion}, depending of where the pure-state excitation is inserted, the correct computation of entanglement entropy requires a careful consideration for all the possible choice of channels. Once again, when the insertion is outside the subsystem, channel 1 (as defined in section \ref{singleinsertion}) is sufficient to determine the full time evolution of entanglement entropy. This is depicted in the left and right graphs of figure \ref{fig:cases}. On the other hand, when the insertion is inside the subsystem, both channel 1 and 2 have to be taken into account in order to compute the correct evolution. Furthermore, we can observe entanglement suppression by fixing the location of the mixed-state operator in spacetime and looking at $\Delta S_A$ as we only let the local operator excitation evolve in time (see figure \ref{fig:MLQ1}).

Finally, we can compute $\Delta S_{A^c}$ and see whether $\Delta S_A = \Delta S_{A^c}$, as is the case for a single local operator (pure-state) excitation. Taking the separation to be lightlike once again, in the limit $a \to -\inf$ and $t \gg b-x_P$ (we are assuming a case II setup here as a complement of the initial case I setup),
\begin{align}
    z &= \frac{\left(e^{-\frac{2\pi^2}{\beta}}-\left(-\frac{b-t+t_M+is}{b-t+t_M-is}\right)^{\frac{2\pi i}{\beta}}\right)\left(\frac{8\pi i\delta}{\beta s}\right)}{\left(e^{-\frac{2\pi^2}{\beta}}-1\right)\left(\left(-\frac{b-t+t_M+is}{b-t+t_M-is}\right)^{\frac{2\pi i}{\beta}}-1\right)} \nonumber\\
    &= \frac{\left(-e^{-\frac{2\pi^2}{\beta}}\frac{4\pi s}{\beta t}\right)\left(\frac{8\pi i\delta}{\beta s}\right)}{\left(e^{-\frac{2\pi^2}{\beta}}-1\right)\left(e^{-\frac{2\pi^2}{\beta}}-1\right)} \nonumber\\
    &= -i\frac{8\pi^2\delta}{\beta^2\sinh^2\left(\frac{\pi^2}{\beta}\right)t} + O(t^{-2}),
\end{align}
or equivalently,
\begin{equation}
    \theta = 2\pi m + \frac{8\pi^2\delta}{\beta^2\sinh^2\left(\frac{\pi^2}{\beta}\right)t} + O(t^{-2}),
\end{equation}
where $m=-1$ or $0$. Similarly for $\bar{z}$, we get
\begin{align}
    \bar{z} &= \frac{\left(e^{-\frac{2\pi^2}{\beta}}-\left(-\frac{-b-t+t_M+is}{-b-t+t_M-is}\right)^{-\frac{2\pi i}{\beta}}\right)\left(-e^{-\frac{2\pi^2}{\beta}}\frac{2\pi i\delta s}{\beta t_M^2}\right)}{\left(-e^{-\frac{2\pi^2}{\beta}}\frac{2\pi s}{\beta t_M}\right)\left(\left(-\frac{-b-t+t_M+is}{-b-t+t_M-is}\right)^{-\frac{2\pi i}{\beta}}-e^{-\frac{2\pi^2}{\beta}}\right)} \nonumber\\
    &= \frac{\left(e^{-\frac{2\pi^2}{\beta}}-e^{\frac{2\pi^2}{\beta}}\right)\left(-e^{-\frac{2\pi^2}{\beta}}\frac{2\pi i\delta s}{\beta t_M^2}\right)}{\left(-e^{-\frac{2\pi^2}{\beta}}\frac{2\pi s}{\beta t_M}\right)\left(e^{\frac{2\pi^2}{\beta}}-e^{-\frac{2\pi^2}{\beta}}\right)} \nonumber\\
    &= -i\frac{\delta}{t_M} + O(t^{-1}),
\end{align}
or equivalently,
\begin{equation}
    \bar{\theta} = 2\pi \bar{m} - \frac{\delta}{t_M} + O(t^{-1}),
\end{equation}
where $\bar{m}=0$ or $1$. The only permissible pairs of $(m,\bar{m})$ are $(0,1)$ and $(-1,0)$, so $\Delta S_{A^c}$ comes out to be
\begin{align}\label{EEpurecomp}
    \Delta S_{A^c} &= \min\left(\frac{c}{6}\log\left(\frac{2t_M}{\alpha_H \delta}\sin(\alpha_H \pi)\right),\frac{c}{6}\log\left(\frac{\beta^2t}{4\pi^2 \alpha_H \delta}\sin(\alpha_H \pi)\sinh^2\left(\frac{\pi^2}{\beta}\right)\right)\right) \nonumber\\
    &= \frac{c}{6}\log\left(\frac{2t_M}{\alpha_H \delta}\sin(\alpha_H \pi)\right).
\end{align}

Comparing (\ref{EEpure}) and (\ref{EEpurecomp}), we see that in general $\Delta S_A \neq \Delta S_{A^c}$, which implies that the doubly-excited state obtained in our setup is not a pure state, but rather a mixed state. This is not necessarily a surprising result considering the insertion of a mixed-state local quench, but the fact that its effect still comes through in $\Delta S_A$ (remembering that this measures the entanglement entropy relative to the `vacuum', which already takes the mixed-state excitation into account) is not at all trivial and is somewhat profound.

\subsection{When $s \ll \delta$}\label{mixedlimit}
\begin{figure}[h]
    \centering
    \includegraphics[width=0.5\linewidth]{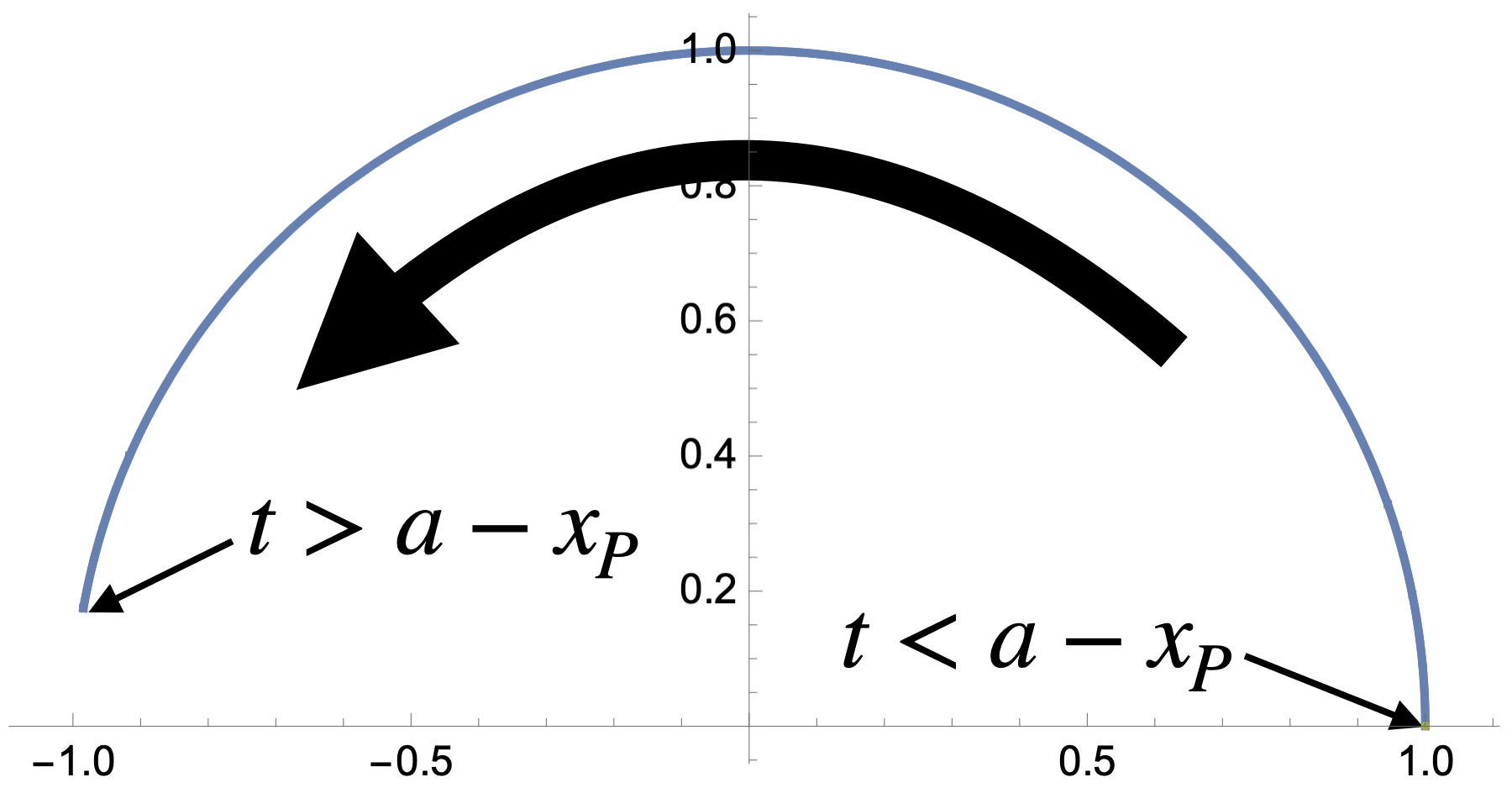}
    \caption{Behavior of $1-z$ as time evolves. In the limit $s \ll \delta$, we only observe an anticlockwise $\pi$ revolution, which once again gives finite entanglement entropy.}
    \label{fig:z_behavior_opplim}
\end{figure}

Let us, once again, first take the separation between excitations to be lightlike, i.e. $x_P+t_M = 0$, and first consider a case I setup. As touched upon in section \ref{4ptfunc}, $\zeta_{\mathcal{O}}, \zeta_{\mathcal{O}^\dagger} \to (-1)^{\frac{2\pi i}\beta}$ in the limit $s \ll \delta$, so we must decide on the logarithmic branch of $-1$. The most natural candidates are $e^{\pm \pi i}$, which give $\zeta_{\mathcal{O}}, \zeta_{\mathcal{O}^\dagger} \to e^{\mp\frac{2\pi^2}\beta}$. Fortunately, when the separation is taken to be lightlike, the choice of branch does not affect the computation, as, in the limit of $b\to\inf$ and $a \ll t$, we obtain
\begin{align}    
    z &= \frac{\left(e^{-\frac{2\pi^2}{\beta}}-e^\frac{2\pi^2}{\beta}\right)\left(e^{-\frac{2\pi^2}{\beta}}\frac{8\pi is}{\beta \delta}\right)}{\left(-e^{-\frac{2\pi^2}{\beta}}\frac{4\pi is}{\beta \delta}\right)\left(e^{\frac{2\pi^2}{\beta}}-e^{-\frac{2\pi^2}{\beta}}\right)} \nonumber\\
    &= 2 + O(t^{-1})
\end{align}
on the $e^{+\pi i}$ branch (which we shall call the `positive branch'), and
\begin{align}    
    z &= \frac{\left(e^{-\frac{2\pi^2}{\beta}}-e^\frac{2\pi^2}{\beta}\right)\left(e^{\frac{2\pi^2}{\beta}}\frac{8\pi is}{\beta \delta}\right)}{\left(e^{-\frac{2\pi^2}{\beta}}-e^{\frac{2\pi^2}{\beta}}\right)\left(e^{\frac{2\pi^2}{\beta}}\frac{4\pi is}{\beta \delta}\right)} \nonumber\\
    &= 2 + O(t^{-1})
\end{align}
on the $e^{-\pi i}$ (`negative') branch, which either way implies a $\pi$ revolution of $1-z$ instead of a $2\pi$ revolution that we saw in the previous section. We can indeed verify that $\theta = \pi$ in figure \ref{fig:z_behavior_opplim}.
$1-\bar{z}$ once again does not exhibit monodromy under time evolution, so (\ref{HHLLsimple}) yields
\begin{equation}
    \Delta S_A = \frac{c}{6}\log \left(\frac{\sin\left(\frac{\alpha_H\pi}{2}\right)}{\alpha_H}\right).
\end{equation}

Although not relevant, we can check that $\bar{z}$ does indeed depend on the choice of branch, as
\begin{align}
    \bar{z} &= \frac{\left(-e^\frac{2\pi^2}{\beta}\frac{4\pi s}{\beta t}\right)\left(-e^{\frac{2\pi^2}{\beta}}\frac{2\pi i\delta s}{\beta t_M^2}\right)}{\left(-e^\frac{2\pi^2}{\beta}\frac{4\pi s}{\beta t}-e^{\frac{2\pi^2}{\beta}}\frac{2\pi s}{\beta t_M}\right)\left(-e^{\frac{2\pi^2}{\beta}}\frac{2\pi s}{\beta t_M}\right)} \nonumber\\
    &= i\frac{2\delta}{t} + O(t^{-2})
\end{align}
on the positive branch, and
\begin{align}
    \bar{z} &= \frac{\left(-e^\frac{2\pi^2}{\beta}\frac{4\pi s}{\beta t}\right)\left(-e^{-\frac{2\pi^2}{\beta}}\frac{2\pi i\delta s}{\beta t_M^2}\right)}{\left(e^\frac{2\pi^2}{\beta}-e^{-\frac{2\pi^2}{\beta}}\right)\left(e^\frac{2\pi^2}{\beta}-e^{-\frac{2\pi^2}{\beta}}\right)} \nonumber\\
    &= i\frac{2\pi^2\delta s^2}{\beta^2 t_M^2 \sinh^2\left(\frac{2\pi^2}{\beta}\right)t} + O(t^{-2})
\end{align}
on the negative branch.

The fact that $1-z$ only makes a $\pi$ revolution has significant implications when we consider a case II setup. When $a<x_P$, $1-z$ loses time dependence and is always $\theta=\pm\pi$ where its sign depends on the channel. This means that it is impossible to set $\theta=0$ or even $|\theta| \ll 1$ for $t<x_P-a$, which is necessary in order to impose $\Delta S_A=0$ at early times. So, necessarily $\Delta S_A \neq 0$ for $t<x_P-a$, which is highly counterintuitive from the point of view of causality.

This conundrum can be partially salvaged by remembering that these local quenches are inserted by hand, so the non-vanishing of $\Delta S_A$ at early times is actually contained within $t_M<t<x_P-a$. At $t<t_M$, the mixed-state local quench has not yet taken place, so we should revert to results from section \ref{Sec:HHLL} and conclude that $\Delta S_A=0$ at early times before $t=t_M$.

The non-vanishing of $\Delta S_A$ for $t_M<t<x_P-a$ remains a mystery. It may be due to the fact that the background on which the local operator quench takes place is a mixed state, and that information about this background somehow spills over to $\Delta S_A$. This is a point that was touched upon in the previous subsection when considering $\delta \ll s$, as we saw that $\Delta S_A \neq \Delta S_{A^c}$. Another possibility is that the cylindrical approximation of the setup geometry --- and therefore the HHLL approximation of conformal blocks --- fails at lightlike separation of excitations. As we will see in the following subsection where we test how duality (\ref{dualityr}) holds up in the setup of a double local quench, we obtain an anomalous term at lightlike separation, which suggests a possible breakdown of the approximation. A further discussion on the validity of the HHLL approximation is provided in section \ref{validity}.

When the separation is non-lightlike, i.e. $|x_P+t_M|\gg\delta,s$, as touched upon earlier, we have to make a choice of branch for $\zeta_\mathcal{O},\zeta_{\mathcal{O}^\dagger}$. Without justification (see the discussion in section \ref{mixedgeneral}), let us take the negative branch for $x_P+t_M>0$ and the positive branch for $x_P+t_M<0$. Somewhat surprisingly, it turns out that
\begin{align}
    z &= \frac{\left(e^{-\frac{2\pi^2}{\beta}}-e^\frac{2\pi^2}{\beta}\right)\left(e^{\frac{2\pi^2}{\beta}}\frac{8\pi i\delta s}{\beta (x_P+t_M)^2}\right)}{\left(e^{-\frac{2\pi^2}{\beta}}-e^{\frac{2\pi^2}{\beta}}\right)\left(e^{\frac{2\pi^2}{\beta}}\frac{4\pi s}{\beta (x_P+t_M)}\right)} \nonumber \\
    &= i\frac{2\delta}{x_P+t_M} + O(t^{-1})
\end{align}
and
\begin{equation}
    \bar{z} = i\frac{8\pi^2\delta s^2}{\beta^2 (-x_P+t_M)^2 \sinh^2\left(\frac{2\pi^2}{\beta}\right)t} + O(t^{-2})
\end{equation}
when $x_P+t_M>0$, and
\begin{align}    
    z &= \frac{\left(e^{-\frac{2\pi^2}{\beta}}-e^\frac{2\pi^2}{\beta}\right)\left(e^{-\frac{2\pi^2}{\beta}}\frac{8\pi i\delta s}{\beta (x_P+t_M)^2}\right)}{\left(e^{-\frac{2\pi^2}{\beta}}\frac{4\pi s}{\beta (x_P+t_M)}\right)\left(e^\frac{2\pi^2}{\beta}-e^{-\frac{2\pi^2}{\beta}}\right)} \nonumber \\
    &= -i\frac{2\delta}{x_P+t_M} + O(t^{-1})
\end{align}
and
\begin{equation}
    \bar{z} = i\frac{2\delta}{t} + O(t^{-2})
\end{equation}
when $x_P+t_M<0$. Altogether, this should give us
\begin{equation}
    \Delta S_A = \frac{c}{6}\log\left(\frac{|x_P+t_M|}{\alpha_H\delta}\sin(\alpha_H \pi)\right)
\end{equation}
--- identical to (\ref{nonlightlike}), which was obtained under the opposite limit of the regulators $\delta \ll s$.

\subsection{Duality of double local quenches}\label{dualitydouble}

Thus far, we have done analytic calculations of $\Delta S_A$ in various limits of the regulators as well as the separation between the quenches. We would like to use this section to briefly summarize the results we have obtained so far and comment on how these results relate to each other through the lens of the duality, proposed in section \ref{dualitysingle}, between pure-state (local operator) and mixed-state quenches.

Taking the subsystem to be semi-infinite ($b\to\infty$), here are the various late-time expressions for $S_A = S_A^\text{mixed} + \Delta S_A$ in the different limits we have looked into:
\begin{itemize}
    \item When $\delta \ll s$
    \begin{itemize}
        \item Lightlike separation ($x_P+t_M=0$)
        \begin{equation}\label{EEA}
            S_A=\frac{c}{3}\log\frac{b}{\ep}+\frac{c}{6}\log\frac{t}{\delta}+\frac{c}{3}\log\left(\frac{\beta}{2\pi}\sinh\left(\frac{\pi^2}{\beta}\right)\right)+\frac{c}{6}\log\left(\frac{\sin(\alpha_H \pi)}{\ap_H}\right)
        \end{equation}
        \item Non-lightlike separation ($|x_P+t_M|\gg \delta,s$)
        \begin{multline}\label{EEB}
            S_A=\frac{c}{3}\log\frac{b}{\ep}+\frac{c}{6}\log\frac{t|x_P+t_M|}{s\delta}\\
            +\frac{c}{6}\log\left(\frac{\beta}{2\pi}\sinh^2\left(\frac{2\pi^2}{\beta}\right)\right)+\frac{c}{6}\log\left(\frac{\sin(\alpha_H \pi)}{\ap_H}\right)
        \end{multline}
    \end{itemize}
    \item When $s \ll \delta$
    \begin{itemize}
        \item Lightlike separation ($x_P+t_M=0$)
        \begin{multline}\label{EEC}
            S_A=\frac{c}{3}\log\frac{b}{\ep}+\frac{c}{6}\log\frac{t}{s}\\
            +\frac{c}{6}\log\left(\frac{\beta}{2\pi}\sinh\left(\frac{2\pi^2}{\beta}\right)\right)+\frac{c}{6}\log\left(\frac{\sin\left(\frac{\alpha_H\pi}{2} \right)}{\ap_H}\right)
        \end{multline}
        \item Non-lightlike separation ($|x_P+t_M|\gg \delta,s$)
        \begin{multline}\label{EED}
            S_A=\frac{c}{3}\log\frac{b}{\ep}+\frac{c}{6}\log\frac{t|x_P+t_M|}{s\delta}\\
            +\frac{c}{6}\log\left(\frac{\beta}{2\pi}\sinh^2\left(\frac{2\pi^2}{\beta}\right)\right)+\frac{c}{6}\log\left(\frac{\sin(\alpha_H \pi)}{\ap_H}\right)
        \end{multline}
    \end{itemize}
\end{itemize}

Should the duality (\ref{dualityr}) hold, a setup considered in the limit $\delta \ll s$ and at separation $(x_P,t_M)$ should be equivalent to that considered in the limit $s \ll \delta$ and at separation $(-x_P,-t_M)$. Since all the expressions above are invariant under the transformation $(x_P,t_M) \mapsto (-x_P,-t_M)$, this means that (\ref{EEA}) should map to (\ref{EEC}) and (\ref{EEB}) to (\ref{EED}) under the swaps of variables
\begin{equation}\label{varswap}
    i\alpha_H \leftrightarrow \frac{2\pi}{\beta}, \qquad \delta \leftrightarrow s.
\end{equation}

Let us start with the more obvious case, when the separation is non-lightlike. The expressions for entanglement entropy (\ref{EEB}) and (\ref{EED}) are identical and invariant under the swaps of variables (\ref{varswap}). Therefore, the duality (\ref{dualityr}) is manifestly satisfied.

Things become more interesting when the separation is lightlike. Under the swaps (\ref{varswap}), all but one of terms in (\ref{EEA}) map to their counterparts in (\ref{EEC}). The anomalous terms that do not map to each other are 
\begin{equation}\label{anomA}
    \frac{c}{3}\log\left(\frac{\beta}{2\pi}\sinh\left(\frac{\pi^2}{\beta}\right)\right)
\end{equation}
from (\ref{EEA}) and
\begin{equation}\label{anomB}
    \frac{c}{6}\log\left(\frac{\sin\left(\frac{\alpha_H\pi}{2}\right)}{\ap_H}\right)
\end{equation}
from (\ref{EEC}); the coefficient of the former is double that of the latter.

We can come up with two possible explanations for the small mismatch between the coefficients of (\ref{anomA}) and (\ref{anomB}). One possibility is simply that our approximation using HHLL conformal blocks breaks down at lightlike separation and so at least one of the results (\ref{EEA}) or (\ref{EEC}) is unreliable. Another possible explanation is that the mixed-state excitation gives an extra thermal contribution to $S_A$ compared to the pure-state excitation, which can be attributed to the fact that entanglement entropy is a fine-grained quantity that can differentiate between pure and mixed states. This was also discussed towards the end of section \ref{dualitysingle}.

\subsection{Energy density}

We shall wrap up this section by computing the energy density of the doubly-excited state. This is motivated by the fact that the energy density is proportional to the entanglement entropy in the limit where the size of subsystem $A$ becomes small, which is a direct consequence of the first law of entanglement entropy \cite{Bhattacharya:2012mi, Blanco:2013joa, Wong:2013gua}. For the sake of convenience, here we will limit our discussion to the case where $\delta \ll s$. One can find that results for $s \ll \delta$ can be obtained in a similar manner once a choice is made on the branch of $\zeta_{\mathcal{O}}, \zeta_{\mathcal{O}^\dagger}$.

The energy density is computed as the sum of the holomorphic and antiholomorphic parts of the stress-energy tensor. For the holomorphic part, we find that at $(X,\bar{X})=(x,x)$,
\begin{align}
    & \frac{\langle T(x,x) O^\dagger(i(\delta-it)+x_P,-i(\delta-it)+x_P) O(-i(\delta+it)+x_P,i(\delta+it)+x_P)\rangle}{\langle O^\dagger(i(\delta-it)+x_P,-i(\delta-it)+x_P) O(-i(\delta+it)+x_P,i(\delta+it)+x_P)\rangle} \nn\\
    & = \frac{16\pi^2 hs^2}{\beta^2 ((x-t+t_M)^2 + s^2)^2} \cdot\frac{\left(-\frac{x-t+t_M+is}{x-t+t_M-is}\right)^{\frac{4i\pi}{\beta}}}{\left(\left(-\frac{x-t+t_M+i s}{x-t+t_M-i s}\right)^{\frac{2i\pi}{\beta}}-\left(-\frac{x_P+t_M+i\delta+is}{x_P+t_M+i\delta-is}\right)^{\frac{2i\pi}{\beta}}\right)^2} \nn\\ 
    & \qquad\qquad\qquad\qquad\cdot\frac{\left(\left(-\frac{x_P+t_M+i\delta+is}{x_P+t_M+i\delta-is}\right)^{\frac{2i\pi}{\beta}}-\left(-\frac{x_P+t_M-i\delta+is}{x_P+t_M-i\delta-is}\right)^{\frac{2i\pi}{\beta}}\right)^2}{\left(\left(-\frac{x-t+t_M+is}{x-t+t_M-is}\right)^{\frac{2i\pi}{\beta}}-\left(-\frac{x_P+t_M-i\delta+is}{x_P+t_M-i\delta-is}\right)^{\frac{2i\pi}{\beta}}\right)^2} \nn\\
    &= \frac{16\pi^2 hs^2}{\beta^2 ((x-t+t_M)^2 + s^2)^2} \frac{\zeta(x)^2(\zeta_{\mathcal{O}^\dagger}-\zeta_{\mathcal{O}})^2}{(\zeta(x)-\zeta_{\mathcal{O}^\dagger})^2(\zeta(x)-\zeta_{\mathcal{O}})^2},
\end{align}
excluding the term associated with the Schwarzian derivative arising from the transformation between $(X,\bar{X})$ and $(\zeta,\bar{\zeta})$. Note that the Schwarzian derivative contribution corresponds to the energy density of the mixed-state excitation. Likewise, the antiholomorphic part of the stress-energy tensor can be computed as follows:
\begin{align}
    & \frac{\langle \bar{T}(x,x) O^\dagger(i(\delta-it)+x_P,-i(\delta-it)+x_P) O(-i(\delta+it)+x_P,i(\delta+it)+x_P)\rangle}{\langle O^\dagger(i(\delta-it)+x_P,-i(\delta-it)+x_P) O(-i(\delta+it)+x_P,i(\delta+it)+x_P)\rangle} \nn\\
    & = \frac{16 \pi ^2 h s^2}{\beta^2 ((x+t-t_M)^2 + s^2)^2} \cdot\frac{\left(-\frac{-x-t+t_M+i s}{-x-t+t_M-i s}\right)^{-\frac{4 i \pi }{\beta }}}{\left(\left(-\frac{-x-t+t_M+i s}{-x-t+t_M-i s}\right)^{-\frac{2 i \pi }{\beta }}-\left(-\frac{-x_P+t_M +i s+i \delta }{-x_P+t_M -i s+i \delta }\right)^{-\frac{2 i \pi }{\beta }}\right)^2} \nn\\
    & \qquad\qquad\qquad\qquad\cdot\frac{\left(\left(-\frac{-x_P+t_M +i s+i \delta }{-x_P+t_M -i s+i \delta }\right)^{-\frac{2 i \pi }{\beta }}-\left(-\frac{-x_P+t_M +i s-i \delta }{-x_P+t_M -i s-i \delta }\right)^{-\frac{2 i \pi }{\beta }}\right)^2}{\left(\left(-\frac{-x-t+t_M+i s}{-x-t+t_M-i s}\right)^{-\frac{2 i \pi }{\beta }}-\left(-\frac{-x_P+t_M +i s-i \delta }{-x_P+t_M -i s-i \delta }\right)^{-\frac{2 i \pi }{\beta }}\right)^2} \nn\\
    &= \frac{16\pi^2 hs^2}{\beta^2 ((x+t-t_M)^2 + s^2)^2} \frac{\bar{\zeta}(x)^2(\bar{\zeta}_{\mathcal{O}^\dagger}-\bar{\zeta}_{\mathcal{O}})^2}{(\bar{\zeta}(x)-\bar{\zeta}_{\mathcal{O}^\dagger})^2(\bar{\zeta}(x)-\bar{\zeta}_{\mathcal{O}})^2}.
\end{align}

\begin{figure}[h]
  \centering
   \includegraphics[width=0.4\linewidth]{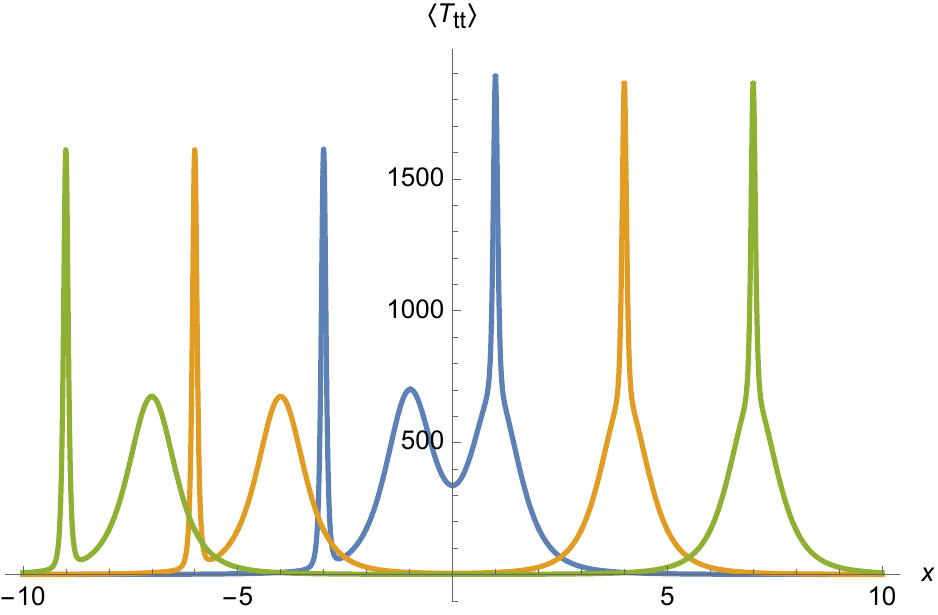}
   \includegraphics[width=0.4\linewidth]{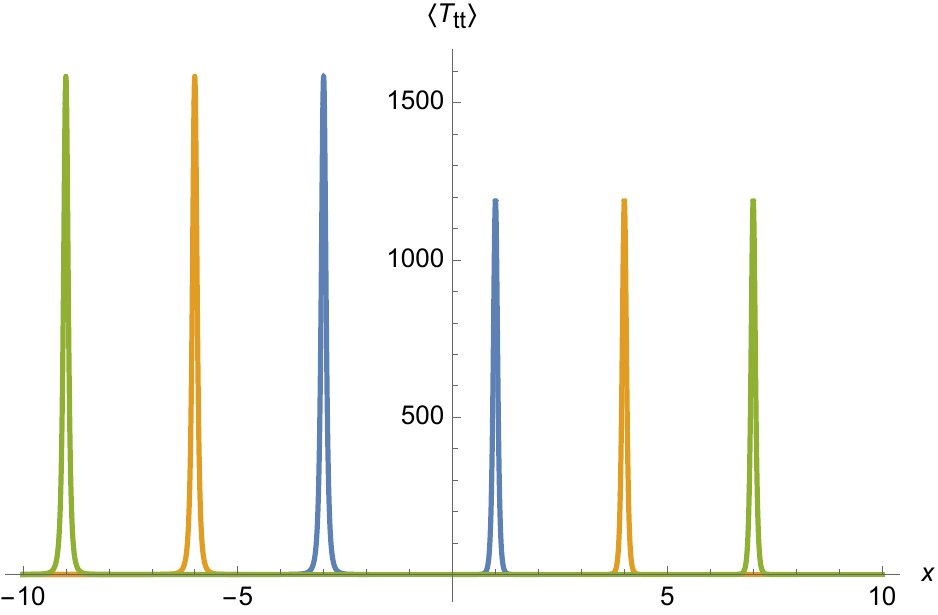}
  \caption{Plots of energy density as a function of position, when the excitations are lightlike-separated ($x_P=-1,t_M=1$) and the regulators are taken to be $\delta=0.1,s=1$. $\beta=1$ (BTZ phase). Time is fixed at $t=2$ (blue), $5$ (orange), and $8$ (green). The left contains contribution from the Schwarzian derivative term, while the right does not. When the regulator is small, the excitation is more locally focused and its contribution to the spike in energy density becomes more pronounced.}
\label{fig:energydensity}
\end{figure}

We can see the general profile of the energy density when $\epsilon \ll s$ in figure \ref{fig:energydensity}. The plot on the left contains the Schwarzian derivative term, which corresponds to the energy contribution due to the mixed-state excitation, while the plot on the right does not. As a check of consistency, we can see that for the left plot we see two peaks in $x<0$ and a single combined peak in $x>0$, which aligns with the fact that $x_P=-1, t_M=1$. For the right plot, the asymmetry in the height of peaks is due to the alignment of the two excitations giving rise to a stronger suppression of entanglement for the right-propagating mode of the entangled pair. The peaks in energy density move away from each other as time evolves, which shows the propagation of modes in opposite directions.

\begin{figure}[h]
  \centering
   \includegraphics[width=0.5\linewidth]{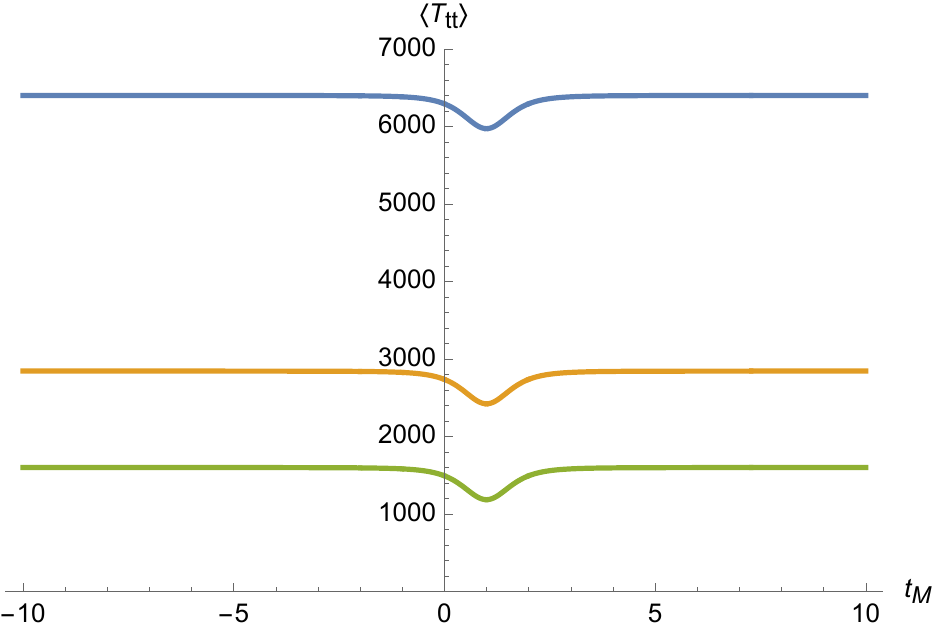}
  \caption{Plots of energy density as a function of $t_M$. Other parameters are fixed at $x=9, t=10, x_P=-1$. The regulator for the mixed-state excitation is set to be $s=1$, while the pure-state regulator is varied between $\delta = 0.05$ (blue), $0.75$ (orange), and $0.1$ (green). $\beta=1$ (BTZ phase) for all plots.} 
\label{fig:energydensity_relpos}
\end{figure}

Figure \ref{fig:energydensity_relpos}, which plots the dependence of energy density on the location of the mixed-state excitation, is also fully consistent with what we expect; energy density is suppressed when $t_M=1$, i.e. when the excitations are lightlike-separated. This provides another qualitative explanation of entanglement suppression shown in figure \ref{fig:caseIImixedfixed}.

\begin{figure}[h]
    \centering
    \includegraphics[width=0.4\linewidth]{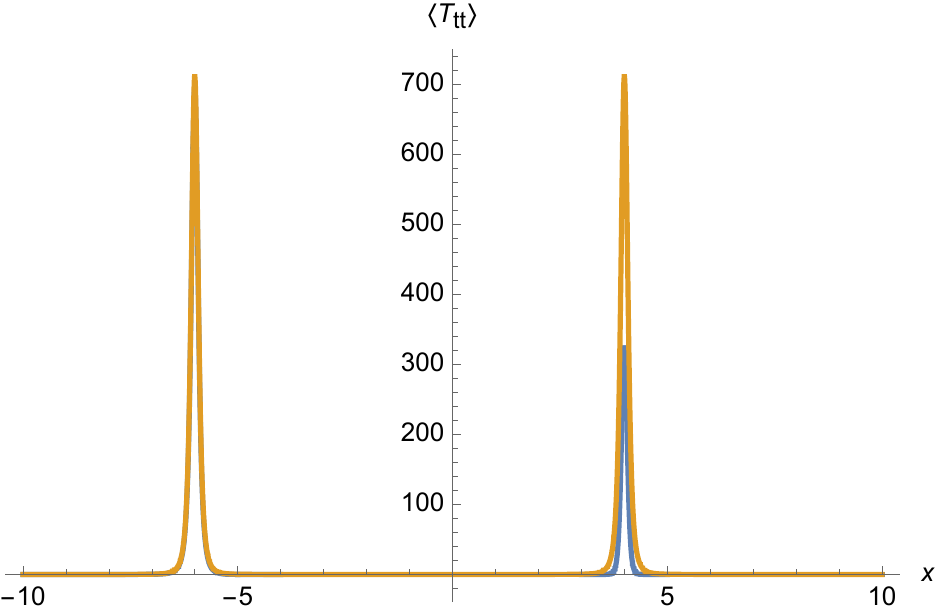}
    \includegraphics[width=0.4\linewidth]{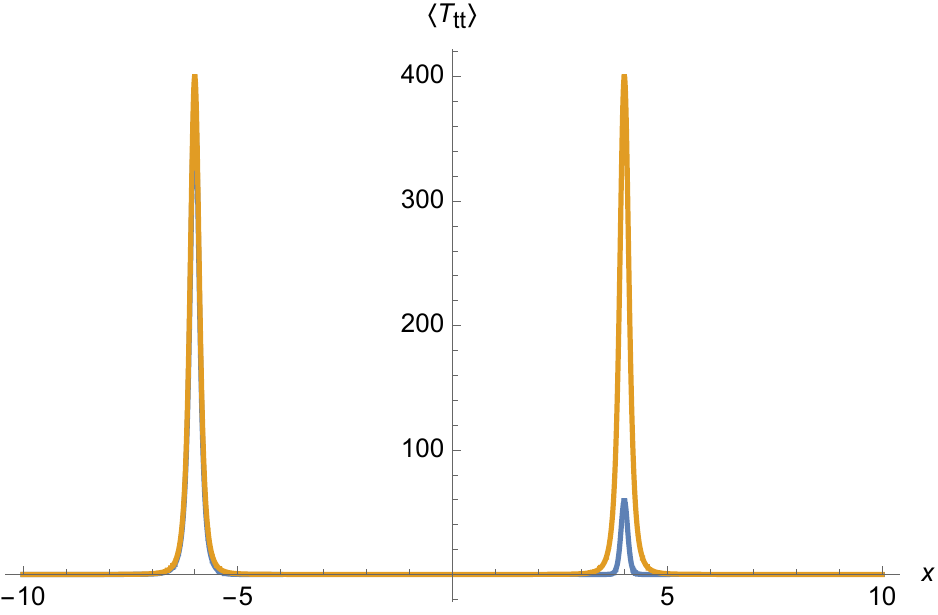}
    \caption{Energy density of a local operator quench with (blue) and without (orange) a mixed-state excitation in the background. The excitations are located at $x_P=-1,t_M=1$. The snapshots are taken at $t=5$ and other fixed parameters are taken to be $s=1,\beta=1$. For the left panel, $\delta=0.15$, and for the right panel, $\delta=0.20$.}
    \label{fig:thermalcomp}
\end{figure}

Finally, we can compare this energy density profile to the case of a single local operator quench (without a mixed-state local quench in the vicinity). As can be seen in figure \ref{fig:thermalcomp}, the energy density peaks corresponding to the left-moving mode coincide completely, whereas for the peaks corresponding to the right-moving mode, the effect of the mixed-state excitation is very evident as the peak is heavily suppressed in its presence.

\section{Analysis of entanglement suppression}\label{Sec:suppression}

In the previous section, where we considered limits $\delta/s\ll 1$ and $\delta/s\gg 1$, we observed that the contribution to entanglement entropy by a local operator excitation is suppressed in the presence of a mixed-state excitation, as it loses its originally logarithmic time-dependence. In arriving at these analytic results, we approximated the torus geometry of the mixed-state excitation with an infinitely long cylinder by picking up the BTZ saddle. To justify this analysis, we need to understand whether this approximation is valid even in the presence of other local operator excitations. This section shall be devoted to discussing this problem.

\subsection{Validity of our approximation}\label{validity}

\begin{figure}[h]
  \centering
   \includegraphics[width=0.4\linewidth]{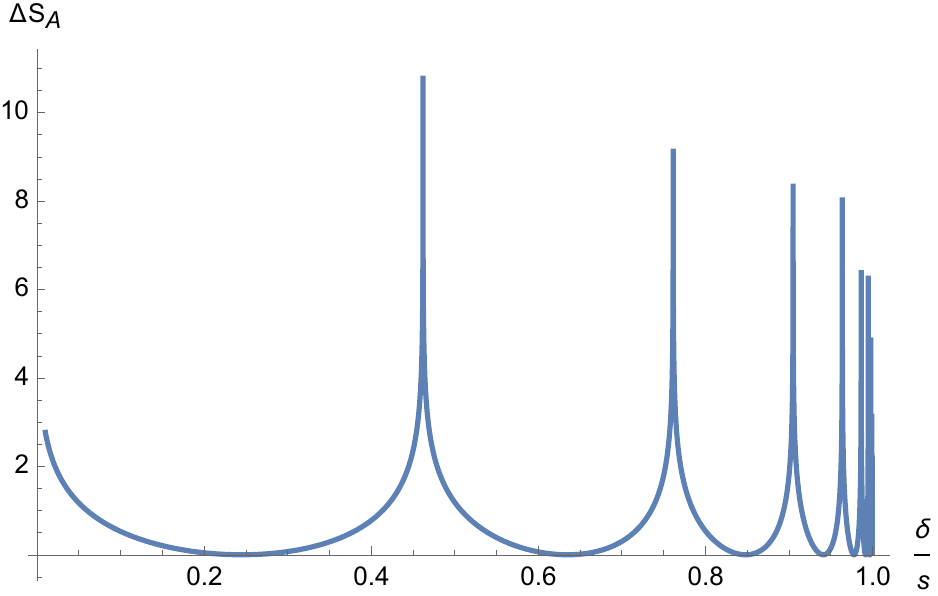}
   \includegraphics[width=0.4\linewidth]{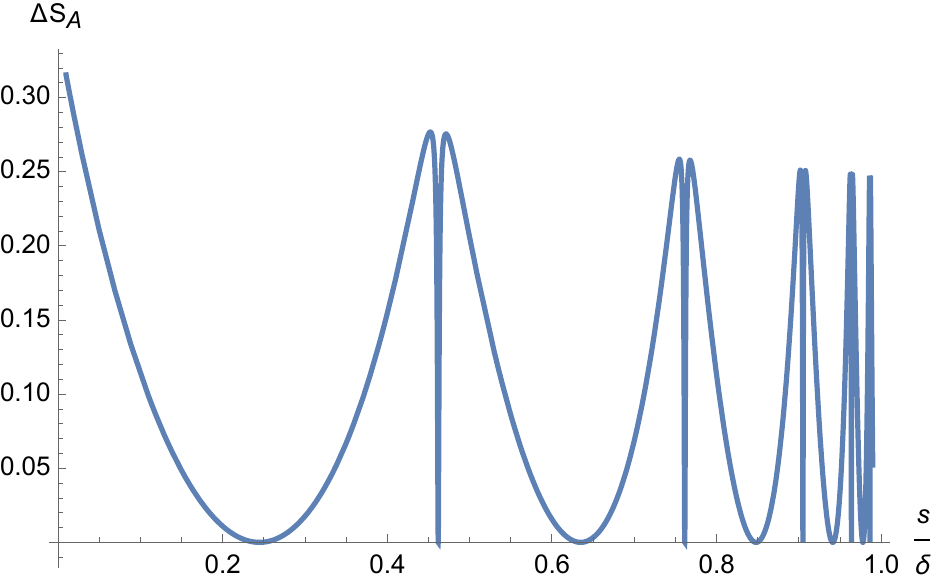}
  \caption{Plots of $\Delta S_A$ as $\delta/s$ is varied. On the left we have $\delta/s < 1$, and on the right we have $\delta/s > 1$. The subsystem is taken to be $a=0,b=1000$ and the separation between excitations is taken to be lightlike and small relative to the subsystem ($x_P=-1,t_M=1$). $\beta=1$ (BTZ phase). These are snapshots of $\Delta S_A$ when $t=500$. Details of computation for $\delta/s > 1$ regarding the choice of branch is discussed in section \ref{mixedgeneral}.} 
\label{fig:EEpeak}
\end{figure}

By employing the approximation of the torus as an infinite cylinder, it is straightforward to evaluate the time evolution of entanglement entropy for any values of $\delta/s$. Here remember that $\delta$ and $s$ play the roles of regularization parameters for the pure-state and mixed-state excitation, respectively. However, as plotted in figure \ref{fig:EEpeak}, when $\delta/s = O(1)$, away from the limits discussed in sections \ref{purelimit} and \ref{mixedlimit}, it turns out that $\Delta S_A$ starts to exhibit singular behavior; it becomes divergent for infinitely many values of $\delta/s$. In the following subsection, we will show that, assuming a lightlike separation between excitations, the divergence occurs when $\delta = \tanh(n\beta/2)s$ for integer $n$.

This singular behavior of $\Delta S_A$ clearly looks unphysical. Moreover, we saw earlier in section \ref{dualitydouble} that, even away from $\delta/s = O(1)$, the analytic expressions obtained in the limits $\delta/s\ll 1$ and $\delta/s\gg 1$ for the total entanglement entropy $S_A$, when the separation is lightlike, do not map to each other under the duality transformation (\ref{varswap}). If we remember how we computed the entanglement entropy, we notice that the only approximation we employed which requires proper justification is to regard correlation functions on a torus as those on an infinite cylinder by picking up the BTZ saddle. Even though this is a quite common approximation in AdS/CFT, we have to be careful now because we inserted a heavy local operator in addition to the mixed-state excitation. The latter is described as a BTZ black hole after a coordinate transformation. The former excitation introduces a time-evolving deficit angle, which produces a huge backreaction on the BTZ geometry. 

Though our HHLL approximation is still able to incorporate perturbative backreactions \cite{Fitzpatrick:2014vua}, we may expect nonperturbative effects such as other saddle contributions, which we did not take into account. This nonperturbative contribution becomes enhanced when the two excitations approach each other. Noting that the two excitations are separated in the original Euclidean geometry by $(X_{\text{sep}},\bar{X}_{\text{sep}})=(i(s-\delta)+x_P+t_M,-i(s-\delta)+x_P-t_M)$, we can reasonably expect nonperturbative interactions between the two excitations to be insignificant when $\delta/s\ll 1$, $\delta/s\gg 1$, and/or $|x_P+t_M|\gg\delta,s$. This heuristic holographic argument is depicted in figure \ref{fig:holsketch}. For a more rigorous justification, we would need to calculate the full correlation function on a torus and compare it to that on an infinite cylinder. Since the former is currently not well-understood, we shall leave this as a future problem.

In the following subsections \ref{puregeneral} and \ref{mixedgeneral}, we discuss in detail the specific peculiarities surrounding the computation of entanglement entropy when we depart from nice limits of the regulators.

\subsection{When $\delta<s$}\label{puregeneral}

Once we depart from the limit $\delta \ll s$, we find that $1-z$ undergoes drastic shifts in its behavior under time evolution. In particular, we find that there are particular ratios of $\delta$ and $s$ at which $\theta$ approaches $\pm2\pi$ and the monodromy of $1-z$ becomes more prominent. Away from these ratios, $\theta$ assumes intermediate values (can even be $0$) and its monodromic behavior is less significant. In order to compute these ratios, as before, let us assume a case I setup.

In (\ref{zanalytic}) (and likewise in (\ref{zbaranalytic})), we have a time-independent term in the numerator,
\begin{equation}
    \zeta_{\mathcal{O}^\dagger\mathcal{O}} \coloneq \left(-\frac{x_P+t_M+i\delta+is}{x_P+t_M+i\delta-is}\right)^{\frac{2\pi i}{\beta}}-\left(-\frac{x_P+t_M-i\delta+is}{x_P+t_M-i\delta-is}\right)^{\frac{2\pi i}{\beta}}.
\end{equation}
It is straightforward to show that this term is purely imaginary, and its value happens to dictate the monodromic behavior of $1-z$. Especially, when the term vanishes, $1-z$ loses its time dependence and $z=0$ for all times. This happens when $\zeta_{\mathcal{O}^\dagger}=\left(-\frac{x_P+t_M+i\delta+is}{x_P+t_M+i\delta-is}\right)^{\frac{2\pi i}{\beta}}$ is real, i.e.
\begin{equation}
    \frac{2\pi}{\beta}\log\left(\left|-\frac{x_P+t_M+i\delta+is}{x_P+t_M+i\delta-is}\right|\right) = \pi n
\end{equation}
for some integer $n$. This further simplifies to
\begin{equation}
    \sqrt{\frac{(x_P+t_M)^2 + (\delta+s)^2}{(x_P+t_M)^2 + (\delta-s)^2}} = e^{n\beta/2}.
\end{equation}
Notice that when we take the separation between excitations to be lightlike
\begin{equation}
    \delta = \tanh\left(\frac{n\beta}{4}\right)s.
\end{equation}

\begin{figure}
    \centering
    \includegraphics[width=0.45\linewidth]{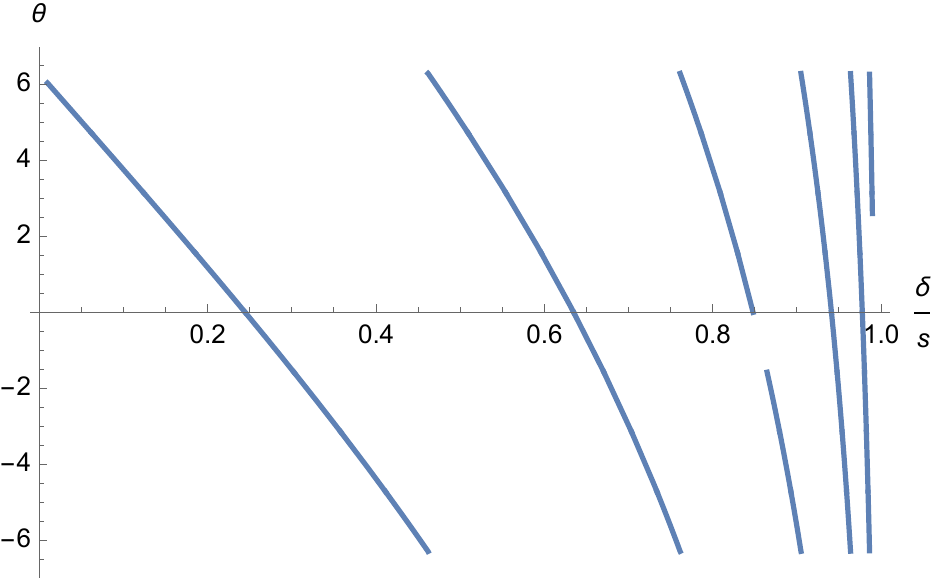}
    \caption{Profile of $\theta$ as a function of $\delta/s$. Here $s$ is fixed at $1$. The monodromic behavior of $1-z$ shifts as the ratio is varied.}
    \label{fig:puremonodromy}
\end{figure}

However, this does not mean that $1-z$ loses its monodromy entirely; this is only case for odd $n$, when $\zeta_{\mathcal{O}^\dagger}$ is negative. For even $n$, $1-z$ makes a full revolution and $\theta = \pm2\pi$ exactly, so $\Delta S_A$ becomes divergent. In fact, by taking the limit $\delta \ll s$ in section \ref{purelimit}, we worked in the vicinity of a divergence in $\Delta S_A$ that occurs at $\delta = 0$. We can therefore conclude that the monodromy of $1-z$ nearing $\theta = \pm 2\pi$ is present only close to
\begin{equation}\label{singularity}
    \delta = \tanh\left(\frac{n\beta}{2}\right)s
\end{equation}
for integer $n$ at lightlike separation. The full dependence of $\theta$ on $\delta/s$ is depicted in figure \ref{fig:puremonodromy}. We can see that the once $\delta$ crosses over $\tanh\left(\frac{n\beta}{4}\right)s$ for odd $n$, the monodromic behavior $(m,\bar{m})\mapsto(m+1,\bar{m})$ (as was the case for $\delta \ll s$) shifts to $(m,\bar{m})\mapsto(m-1,\bar{m})$. We should also note that away from (\ref{singularity}), $|\theta-2\pi m|\ll1$ can no longer be satisfied, so, as commented in section \ref{mixedlimit}, $\Delta S_A = 0$ at $t=0$ is unattainable for case II setups.

\subsection{When $s < \delta$}\label{mixedgeneral}

The argument from the previous subsection also applies (at least qualitatively) when $s < \delta$, and one can check (and see in figure \ref{fig:EEpeak}) that the monodromic behavior of $1-z$ is highly dependent on the ratio of the regulators. However, in this section, we shall discuss another issue that is unique to this regime.

\begin{figure}[h]
  \centering
   \includegraphics[width=0.4\linewidth]{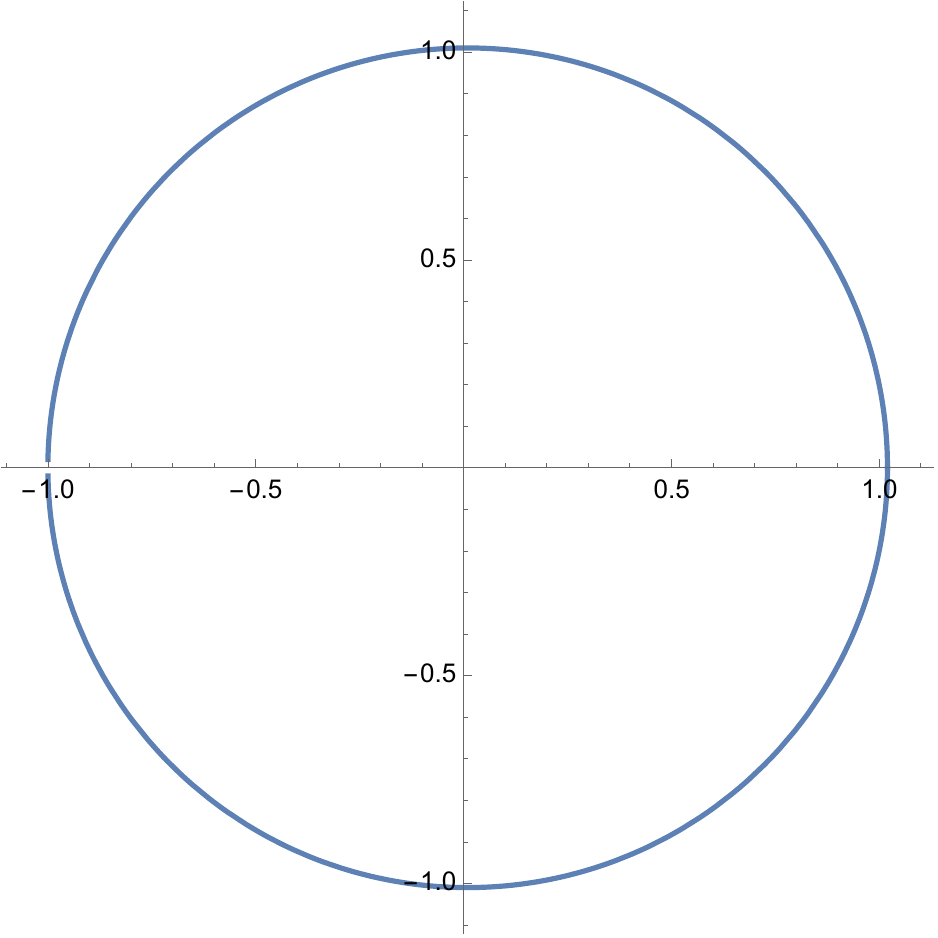}
   \includegraphics[width=0.4\linewidth]{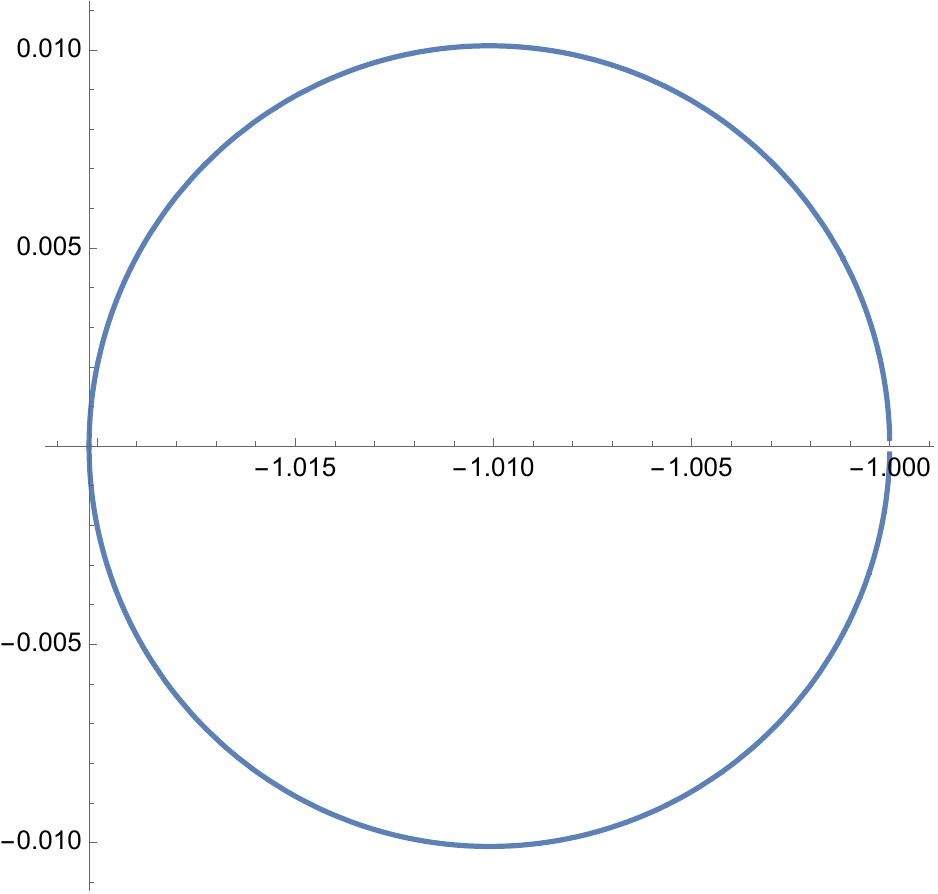}
  \caption{Contours of $\zeta_{\mathcal{O}^\dagger}^{\beta/2\pi i} = e^{w_{\mathcal{O}^\dagger}}$ as $\delta$ is varied, when $\delta \ll s$ (left) and $s \ll \delta$ (right).} 
\label{fig:zetabehavior}
\end{figure}

As alluded to at the end of section \ref{4ptfunc}, the computation of $\Delta S_A$ is less straightforward when $s < \delta$ due to an ambiguity that appears in the calculation of the cross ratios. One might ask why we get this ambiguity only when $s<\delta$ and not when $\delta<s$. The answer to this is well-illustrated in the behavior of $\zeta_{\mathcal{O}}$, $\zeta_{\mathcal{O}^\dagger}$, which drastically changes once we cross over to the $s<\delta$ regime. As can be seen in figure \ref{fig:zetabehavior}, while $\delta<s$, $\zeta_{\mathcal{O}^\dagger}^{\beta/2\pi i} = e^{w_{\mathcal{O}^\dagger}} = -\frac{x_P+t_M+i\delta+is}{x_P+t_M+i\delta-is}$ revolves around the origin as we vary $x_P+t_M$ from $-\inf$ to $\inf$. This gives $\zeta_{\mathcal{O}^\dagger}^{\beta/2\pi i}$ a natural choice of branch going from $e^{\pi i}$ as $x_P+t_M \to -\inf$ to $e^{-\pi i}$ as $x_P+t_M \to \inf$. That is no longer the case when $s < \delta$. It starts at $\zeta_{\mathcal{O}^\dagger}^{\beta/2\pi i} = -1$ but does not go around the origin; instead, it forms a circle in $\Re\left(\zeta_{\mathcal{O}^\dagger}^{\beta/2\pi i}\right)<-1$, so there is seemingly no natural choice of branch for $\zeta_{\mathcal{O}^\dagger}^{\beta/2\pi i}$ when $s<\delta$.

\begin{figure}[h]
  \centering
   \includegraphics[width=0.4\linewidth]{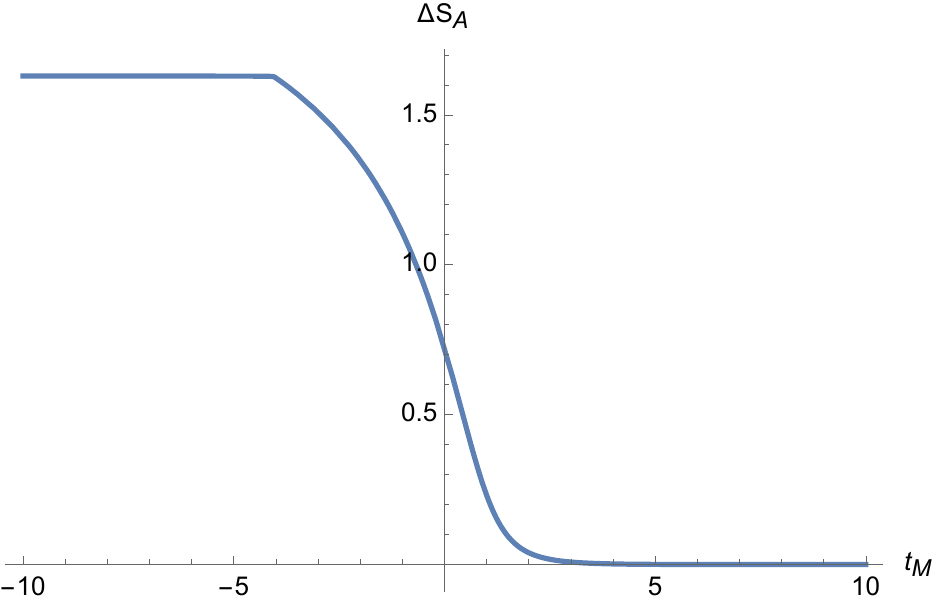}
   \includegraphics[width=0.4\linewidth]{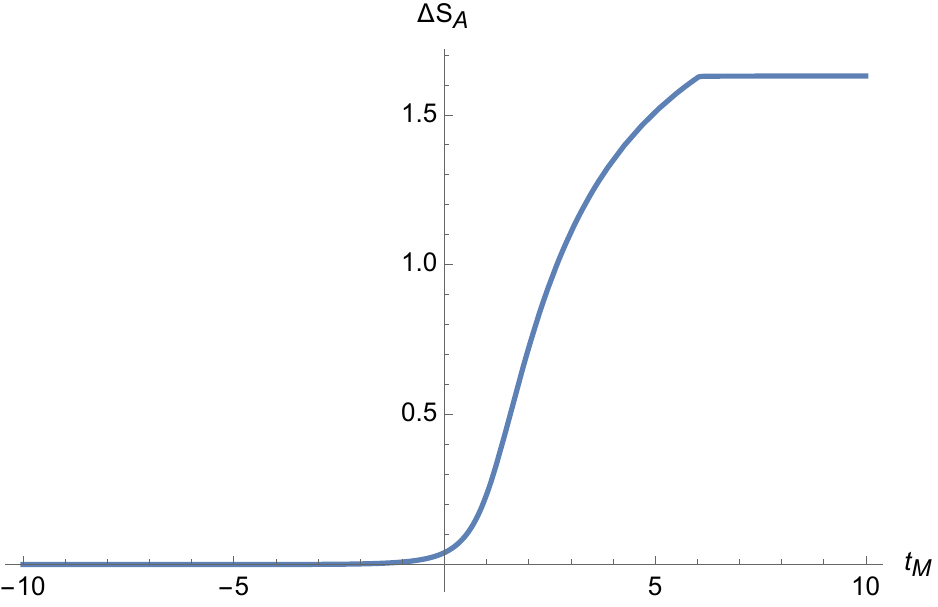}
   \includegraphics[width=0.4\linewidth]{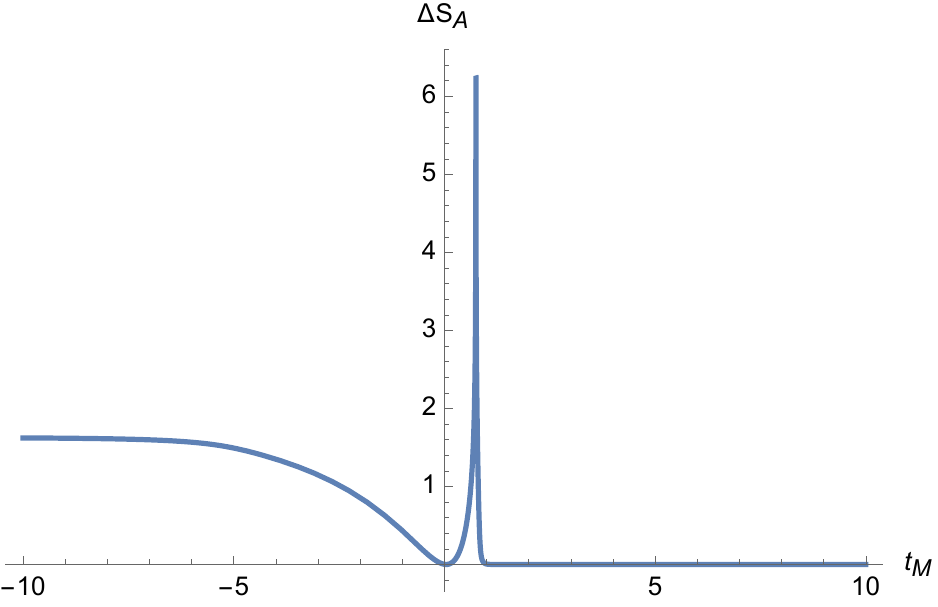}
   \includegraphics[width=0.4\linewidth]{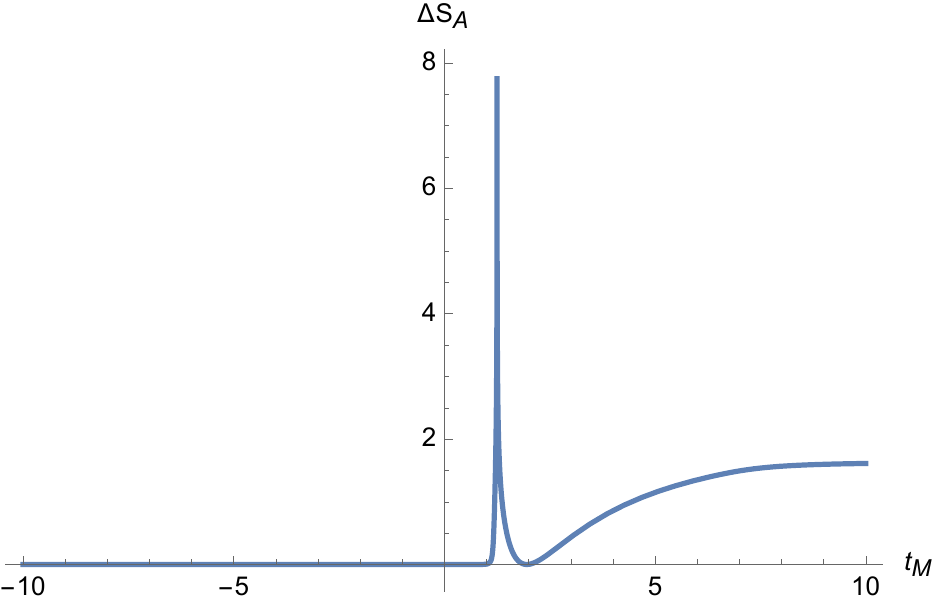}
   \includegraphics[width=0.4\linewidth]{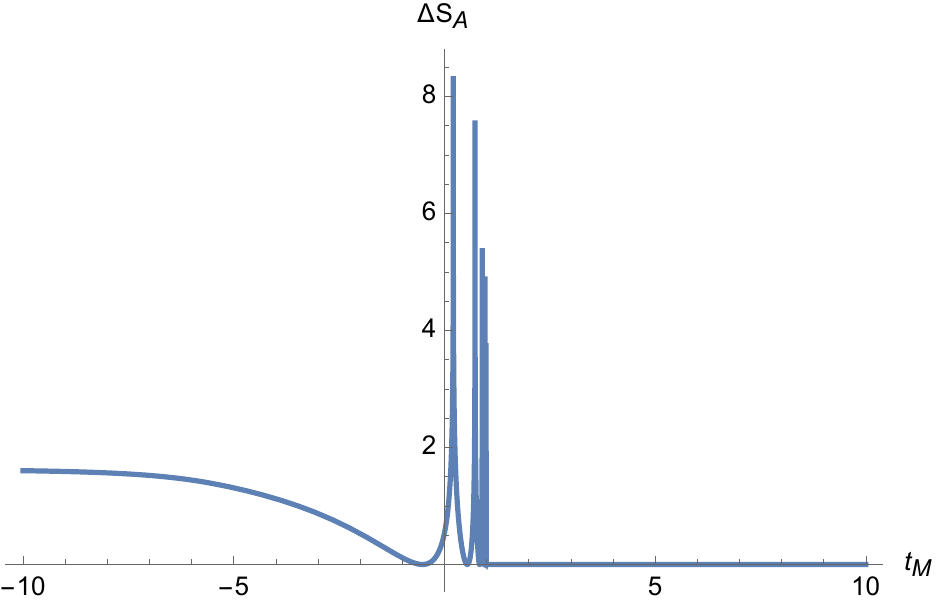}
   \includegraphics[width=0.4\linewidth]{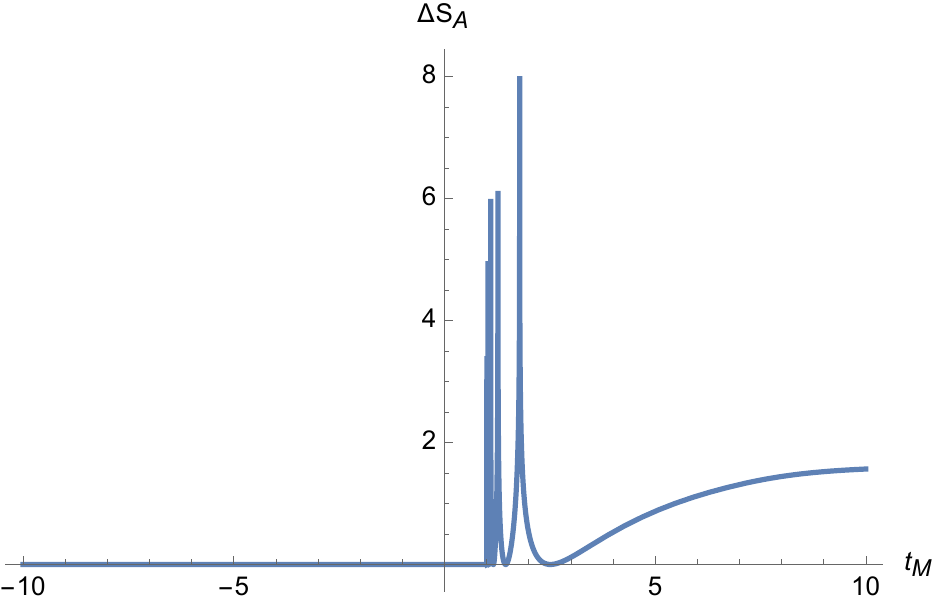}
  \caption{Plots of $\Delta S_A$ as $t_M$ is varied, when the positive branch is chosen (left) and the negative branch is chosen (right). The subsystem is taken to be $a=10,b=20$ and the operator is inserted at $x_P=-1$. These snapshots of $\Delta S_A$ are taken at $t=16$. The ratios of the regulators are chosen to be $s/\delta=0.01,0.50,0.99$ going from top to bottom, where $\delta$ is fixed to be $1$. $\beta=1$ (BTZ phase) for all plots.} 
\label{fig:EEbranches}
\end{figure}

Two plausible choices of branch are $\zeta_{\mathcal{O}^\dagger}^{\beta/2\pi i} = e^{\pi i}$ as $x_P+t_M \to \pm\inf$ (we shall call this the `positive branch') and $\zeta_{\mathcal{O}^\dagger}^{\beta/2\pi i} = e^{-\pi i}$ as $x_P+t_M \to \pm\inf$ (`negative branch'). These choices are compared in figure \ref{fig:EEbranches}, which plots $\Delta S_A$ with respect to $t_M$ while fixing $x_P=-1$. It is immediately clear that these plots are symmetric about $x_P+t_M=0$ ($t_M=1$). The positive branch suppresses contribution to entanglement entropy when $x_P+t_M>0$ and vice versa. As the ratio $s/\delta$ approaches $1$, cutoff at $x_P+t_M=0$ becomes more pronounced and we start seeing singular behavior near the cutoff, which is another reason to doubt the validity of our results when $s/\delta = O(1)$.

Due to the symmetry between these two choices, it is not clear whether either one of these is the correct choice of branch for $\zeta_{\mathcal{O}^\dagger}^{\beta/2\pi i}$. From the viewpoint of the duality (\ref{dualityr}) between pure-state and mixed-state quenches and based on our previous discussion on this in section \ref{dualitydouble}, it is tempting to say that the correct choice is the negative branch for $x_P+t_M>0$ and the positive branch for $x_P+t_M<0$. This, however, means that the loci of $\zeta_\mathcal{O},\zeta_{\mathcal{O}^\dagger}$ as we vary $x_P+t_M$ are discontinuous at $x_P+t_M=0$. Whether this discontinuity is admissible or not is again not immediately clear based on our work so far. Perhaps, this ambiguity could instead reflect a fundamental flaw in our method when employed in the $s<\delta$ regime, which could explain the breakdown of duality in section \ref{dualitydouble}. However, as suggested by the holographic discussion in sections \ref{validity} and \ref{holoint}, there is seemingly no reason to believe a priori that our method would fail even in this regime, especially when $s\ll\delta$. A complete understanding of this ambiguity will require further work.

\subsection{Holographic interpretation}\label{holoint}

Finally, we would like to examine the issue of the
$\delta/s=O(1)$ regime from a holographic perspective, following a similar argument in section \ref{sec:holg} for the (single) local operator quench.

We shall once again start from the original two-hole geometry with coordinates $(X,\bar{X})$, as depicted in figure \ref{fig:setup}. The physical UV cutoff (lattice spacing) $\ep$ is introduced in these coordinates. As before, we can conformally map this geometry onto the $(\zeta,\bar{\zeta})$-coordinates (\ref{zetamapp}), where we can approximate the mapped geometry by the standard Euclidean plane with local operator insertions at $\zeta=\zeta_3$ and $\zeta=\zeta_4$. Subsystem $A$ is mapped onto the interval $[\zeta_1,\zeta_2]$. The entanglement entropy reads
\ba
S_A=\frac{c}{6}\log\frac{|\zeta_2-\zeta_1|^2}{\ep^2|\zeta'_1||\zeta'_2|}
+\Delta S_A,  \label{SSSq}
\ea
where the first term is the entanglement entropy of the mixed-state quench and the second term $\Delta S_A$ is the increase in entanglement entropy due to the local operator quench. 

Now, take another conformal map
\ba
z=\frac{(\zeta_1-\zeta)(\zeta_3-\zeta_4)}{(\zeta_1-\zeta_3)(\zeta-\zeta_4)}.
\ea
This maps the local operator insertions to $z_3=1$ and $z_4=\infty$, while the two endpoints of $A$ are mapped to $z_1=0$ and $z_2$. The entanglement entropy now reads
\ba
S_A=\frac{c}{6}\log\frac{|z_2|^2}{\ep^2|z'_1||z'_2|}+\Delta S_A,
\ea
where $\Delta S_A$ is the same as that in (\ref{SSSq}).

Now, we can perform a final conformal mapping of 
\ba
1-z=e^{iy}, \qquad 1-\bar{z}=e^{-i\bar{y}}.
\ea
If we set $(y_2,\bar{y}_2)=(\theta,\bar{\theta})$, then this coincides with (\ref{zphasew}). Following a similar discussion in section \ref{sec:holg}, the gravitational dual of the $(y,\bar{y})$-coordinates can be identified, as the CFT now lives on a cylinder and the local operators are inserted at the cylinder's past and future infinities. Thus, by using the variables introduced in (\ref{thephs}), the entanglement entropy is given by
\ba
S_A=\frac{c}{6}\ti{\rho}_\infty+\frac{c}{6}\log\left(\frac{\sin \left(\frac{\ap}{2}\left(\Delta \ti{\phi}+\Delta \ti{t}\right)\right)\sin \left(\frac{\ap}{2}\left(\Delta \ti{\phi}-\Delta \ti{t}\right)\right)}{\ap^2}\right).
\ea
Here, we employed the well-known fact that the insertion of local operator with dimension $h$ gives rise to a deficit angle of $2\pi(1-\ap)$, where $\ap=\s{1-24h/c}$.

Now let us examine how the studied behavior of the double local quench is reflected in the gravitational dual in terms of the behavior of $\Delta \ti\phi$ and $\Delta \ti t$. Here, we shall limit our discussion to the case I setup. When $\delta/s\ll 1$, we have $0<\theta<2\pi$ and $0<\bar{\theta}\ll1$, which corresponds to $0<-\Delta\ti{t}<\Delta\ti{\phi}<\pi$. Therefore, $\theta\bar{\theta}>0$ and subsystem $A$ in the $(y,\bar{y})$-coordinates is spacelike as expected; the gravitational picture makes sense and perfectly agrees with our prescription for numerical calculations. The same holds when $\delta/s\gg 1$. However, when $\delta/s=O(1)$, we find that subsystem $A$ becomes timelike as $\theta$ starts to assume negative values. This implies that we can no longer trust the gravitational picture described above that involves the deficit angle, or equally, the HHLL approximation. To remedy this situation, we will probably need to treat the two-point function on a torus carefully without any approximations, which we shall leave as future work.

\section{Conclusion and discussion}\label{Sec:CD}

In this paper, we studied the behavior of the time evolution of entanglement entropy of two-dimensional holographic CFTs in the presence of two local excitations. One of them is a pure-state excitation with regulator $\delta$. The other is a localized mixed-state excitation with regulator $s$, which can be thought of as a localized region of size $s$ at high temperature. In the replica method calculation of two-dimensional CFTs, this model has an advantage that we can treat the mixed-state excitation by gluing two holes in the Euclidean path integral, which, via an appropriate conformal map, is equivalent to a path integral on a torus. Therefore, the entanglement entropy for our doubly-excited state can be obtained by computing the four-point function of two twist operators and two local operators on a torus.

Since correlation functions on a torus are difficult to compute in general, we took advantage of the property of holographic CFTs and focused on the higher-temperature (BTZ black hole) saddle point of the mixed-state excitation. By employing the HHLL approximation of the identity conformal block to treat the heavy local operators and light twist operators in the von Neumann limit $n\to 1$, we managed to find an analytic expression of the entanglement entropy for subsystem $A$ that is taken to be an interval.

In a double local quench, the pure-state (local operator) and mixed-state quenches are both obvious sources of entanglement entropy growth. Earlier works showed that each of these two different excitations alone contributes to the logarithmic growth of entanglement entropy $S_A\sim\frac{c}{6}\log t$. While we can naively expect the combination of these excitations to give rise to a time profile of entanglement entropy $S_A\sim \frac{c}{3}\log t$ via a simple sum, we instead found that its growth is half of that expectation, $S_A\sim \frac{c}{6}\log t$. In other words, the contribution of the local operator excitation appears to be absorbed by the mixed-state excitation. 

We argued that this phenomenon, which we call `entanglement suppression', is naturally understood in both CFT and gravitational pictures. On the CFT side, consider the propagation of an entangled pair of quantum modes created by the local operator. These modes travel in opposite directions at the speed of light. This excitation exists in the backdrop of a mixed-state quench, which looks like a locally high-temperature plasma. If we choose the parameters so that one of the entangled modes collides with the mixed-state excitation, then it gets scattered off and becomes disentangled. In other words, it begins to be entangled with the purifier of the mixed-state excitation. Therefore its contribution to $S_A$ becomes suppressed. On the gravity side of AdS$_3/$CFT$_2$, the local operator excitation is dual to a massive particle, which is a time-evolving deficit angle. This produces shock waves which propagate in the bulk AdS. Eventually some of them will travel to subsystem $A$ on the AdS boundary, which contributes to the holographic entanglement entropy. The mixed-state excitation, on the other hand, is dual to a localized black hole in AdS$_3$. When shock waves from the moving deficit angle pass near the localized black hole, some of them will be absorbed into the black hole, thus explaining the mechanism of entanglement suppression. 

However, we found that our analysis works only when $\delta/s\ll 1$ or $\delta/s\gg 1$. When $\delta/s = O(1)$, we observed singular behavior where the growth of entanglement entropy $\Delta S_A$ becomes divergent. We argued that this problem occurs because we approximated the correlation functions on a torus as those on an infinite cylinder. Even though our calculations using HHLL conformal blocks take into account perturbatively the effect of backreactions due to the heavy local operator excitation, it is possible that saddles other than the BTZ black hole may contribute. In addition, we found that the proposed duality between local operator excitations and mixed-state excitations breaks down --- even away from $\delta/s=O(1)$ --- when the separation between excitations is taken to be lightlike, which might be due either to the failure of our HHLL approximation or to an unaccounted-for thermal entropy contribution. To investigate this problem at $\delta/s=O(1)$ and for lightlike separations more rigorously, we need to explicitly evaluate the correlation function on a torus. We leave this as a major future problem.

Our analysis of entanglement entropy of doubly-excited states provides a new approach to the scattering problem of strongly-coupled quantum field theories. Our entanglement suppression is due to strong interactions in two-dimensional holographic CFTs. On the other hand, it is known that in integrable CFTs, entanglement suppression does not occur. In this way, entanglement suppression is expected to be a new probe of the chaotic dynamics in quantum field theories or even in more general quantum many-body systems. It would be intriguing to investigate entanglement suppression in various strongly-correlated quantum many-body systems. At the same time, it would also be intriguing to understand the basic mechanism of why entanglement suppression occurs from the viewpoint of quantum many-body dynamics. One useful approach is to use the connection between holographic spacetimes and tensor networks \cite{Swingle:2009bg,Nozaki:2012zj,Qi:2013caa,Pastawski:2015qua,Hayden:2016cfa,Milsted:2018san}. In particular, the idea of path integral optimization \cite{Caputa:2017urj,Caputa:2017yrh} will provide a direct connection between our Euclidean path integral construction and its tensor network description. We hope to come back to this problem in the near future.

\section*{Acknowledgement}

We are grateful to Pawel Caputa, Veronika Hubeny, Masamichi Miyaji and Mukund Rangamani for useful discussions. We also thank the organizers and participants of the workshop YITP-W-25-06 ``New advancements on defects and their applications'' for the valuable opportunity to present and to receive feedback on this work. This work is supported by by MEXT KAKENHI Grant-in-Aid for Transformative Research Areas (A) through the ``Extreme Universe'' collaboration: Grant Number 21H05187. TT is also supported by Inamori Research Institute for Science and by JSPS Grant-in-Aid for Scientific Research (B) No.~25K01000. KD is supported by JSPS KAKENHI Grant Number JP24KJ1466.

\appendix

\section{Analysis in the thermal AdS phase}\label{ap:thermal}
In the case of thermal AdS ($\beta>2\pi$), the two-hole Euclidean geometry induced by the mixed-state excitation turns out not to be interesting at all. Working with a holographic CFT reduces the geometry back to a full Euclidean plane with no holes, as the temporal direction is decompactified. The HHLL approximation can be carried out in the original geometry $(X,\bar{X})$, and the expression for $\Delta S_A$ is therefore identical to the one computed in section \ref{HHLL}, where we considered a single insertion of a (pure-state) local operator.

A cautious reader may want to verify this result by applying the map to the annulus $X \mapsto e^w$. The map itself does depend on $t_M,s$, which are parameters associated with the mixed-state excitation. However, the cross ratios $z,\bar{z}$ turn out to be independent of either one of these parameters.
\begin{align}
    1-z &= \frac{(a-x_P -t+i \delta ) (b-x_P -t-i \delta )}{(a-x_P -t-i \delta ) (b-x_P -t+i \delta )} \\
    1-\bar{z} &= \frac{(a-x_P +t-i \delta ) (b-x_P +t+i \delta )}{(a-x_P +t+i \delta ) (b-x_P +t-i \delta )}
\end{align}
This agrees with the expression obtained in section \ref{singleinsertion}.

\section{With chemical potential}\label{ap:pauli}

As discussed in section \ref{mixedsingle}, without chemical potential, the BTZ phase was represented as a cylinder with a compactified temporal direction $w \sim w + \beta$. If we add a chemical potential term to the excitation (\ref{mixedexcitation}), the cylinder becomes contorted and the holographic dual to this geometry should be a rotating BTZ black hole. With this in mind, the modification is simple: $w$ should now be identified as $w \sim w + \beta(1-i\Omega)$, so our map to the annulus $\zeta$ should be expressed as
\begin{equation}
    \zeta = e^{\frac{2\pi i}{\beta(1-i\Omega)}w}.
\end{equation}

As usual, we can treat the annulus as a full Euclidean plane, on which we know the standard expression for the two-point function. Transforming back to the cylinder, we obtain
\begin{align}
    \langle\sigma_n(i\theta_a)\bar{\sigma}_n(i\theta_b)\rangle &= \left|\frac{d\zeta}{dw}\right|_{w=i\theta_a}^{\frac{c}{12}(n-\frac{1}{n})}\left|\frac{d\zeta}{dw}\right|_{w=i\theta_b}^{\frac{c}{12}(n-\frac{1}{n})}\langle\sigma_n(\zeta(i\theta_a))\bar{\sigma}_n(\zeta(i\theta_b))\rangle \nonumber\\
    &
    \begin{multlined}
        =\biggl(\frac{\beta^2(1+\Omega^2)}{4\pi^2}e^{\frac{\pi(\theta_a+\theta_b)}{\beta(1-i\Omega)}}e^{\frac{\pi(\bar{\theta}_a+\bar{\theta}_b)}{\beta(1+i\Omega)}} \\
        \quad \frac{(e^{-\frac{2\pi\theta_b}{\beta(1-i\Omega)}}-e^{-\frac{2\pi\theta_a}{\beta(1-i\Omega)}})(e^{-\frac{2\pi\bar{\theta}_b}{\beta(1+i\Omega)}}-e^{-\frac{2\pi\bar{\theta}_a}{\beta(1+i\Omega)}})}{\epsilon^2}\biggr)^{-\frac{c}{12}(n-\frac{1}{n})}
    \end{multlined} \nonumber\\
    &=\left(\frac{\beta^2(1+\Omega^2)}{\pi^2\epsilon^2}\sinh\left(\frac{\pi(\theta_b-\theta_a)}{\beta(1-i\Omega)}\right)\sinh\left(\frac{\pi(\bar{\theta}_b-\bar{\theta}_a)}{\beta(1+i\Omega)}\right)\right)^{-\frac{c}{12}(n-\frac{1}{n})},
\end{align}
which is a natural generalization of (\ref{finitetemp}). If we then apply (\ref{striptotwohole}), and work once again in the limit of a large subsystem $a \ll b$ and in the mid-time regime $0 < r,l \ll a \ll t \ll b$, we find the two-point function on the original two-hole geometry to be
\begin{equation}
    \langle\sigma_n(a)\bar{\sigma}_n(b)\rangle = \left(\frac{\beta(1-i\Omega) b^2 t}{2\pi\epsilon^2\sqrt{r^2-l^2}}\sinh\left(\frac{2\pi^2}{\beta(1-i\Omega)}\right)\right)^{-\frac{c}{12}(n-\frac{1}{n})}.
\end{equation}
The application of the replica trick yields
\begin{equation}
    S_A = \frac{c}{6}\log\left(\frac{\beta(1-i\Omega) b^2 t}{2\pi\epsilon^2\sqrt{r^2-l^2}}\sinh\left(\frac{2\pi^2}{\beta(1-i\Omega)}\right)\right).
\end{equation}

\clearpage
\bibliographystyle{JHEP}
\bibliography{Mixed}


\end{document}